\documentclass{article}

\usepackage{PRIMEarxiv}

\usepackage[utf8]{inputenc} % allow utf-8 input
\usepackage[T1]{fontenc}    % use 8-bit T1 fonts
\usepackage{hyperref}       % hyperlinks
\usepackage{url}            % simple URL typesetting
\usepackage{booktabs}       % professional-quality tables
\usepackage{amsfonts}       % blackboard math symbols
\usepackage{nicefrac}       % compact symbols for 1/2, etc.
\usepackage{microtype}      % microtypography
\usepackage{lipsum}
\usepackage{fancyhdr}       % header
\usepackage{graphicx}       % graphics
\graphicspath{{media/}}     % organize your images and other figures under media/ folder

\usepackage{subcaption}
\usepackage[super]{nth}
\usepackage{amsmath,esint}
\usepackage[export]{adjustbox}
\usepackage{ragged2e}
\usepackage[T1]{fontenc}

\title{A STUDY ON CONVOLUTION NEURAL NETWORK FOR RECONSTRUCTING THE TEMPERATURE FIELD OF WALL-BOUNDED FLOWS}

\author{
  Victor Coppo Leite, Elia Merzari \\
  Ken and Mary Alice Lindquist Department of Nuclear Engineering \\
  Pennsylvania State University \\
  205 Hallowell Bldg, University Park, PA 16802\\
  \texttt{vbc5085\@psu.edu}, \texttt{ebm5351\@psu.edu}\\
   \And
  Roberto Ponciroli, Lander Ibarra \\
  Nuclear Science and Engineering Division \\
  Argonne National Laboratory \\
  Lemont, IL 60439\\
  \texttt{rponciroli\@anl.gov}, \texttt{libarra\@anl.gov} \\
}

\begin{document}
\maketitle

\begin{abstract}

In the present study, the capabilities of a new Convolutional Neural Network (CNN) model are explored with the paramount objective of reconstructing the temperature field of wall-bounded flows based on a limited set of measurement points taken at the boundaries of the fluid domain. For that, we employ an algorithm that leverages the CNN capabilities provided with additional information about the governing equations of the physical problem. Great progress has been made in the recent years towards reconstructing and characterizing the spatial distribution of physical variables of interest using CNNs. In principle, CNNs can represent any continuous mathematical function with a relatively reduced number of parameters. However, depending on the complexity imposed by the physical problem, this technique becomes unfeasible. The present study employs a Physics Informed Neuron Network technique featuring a data-efficient spatial function approximator. As a first proof of concept, the CNN is trained to retrieve the temperature of a heated channel based on a limited number of sensors placed only at the boundaries of the domain. In this context, the training data are the temperature fields solutions considering various flows conditions at steady state, e.g. varying the Reynolds and the Prandtl numbers. Additionally, a demonstration case considering the more complex geometry of a MSR is also provided.

Assessment on the performance of the CNN is done by the mean (\(\bar{L_2}\)) and the maximum (\(L_\infty\)) Euclidean norms stemmed from the difference between the actual solutions and the predictions made by the CNN. Finally, a sensitivity analysis is carried out such that the robustness of the CNN is tested considering a potential real application scenario where noise is inevitable. For that, the original test inputs are overlaid with a normal distribution of random numbers targeting to mimic different levels of noise in the measurement points.

\end{abstract}

% keywords can be removed
\keywords{Data-driven \and Scientific Computing\and Machine Learning \and Field Reconstruction}

\section{Introduction}

The objective of the present study is to test an algorithm that leverages Convolutional Neural Networks (CNNs) to reconstruct the entire temperature field based on local and sparse measurements taken only at the boundaries of a fluid domain. The proposed approach features a completely data-driven deep neural network that is trained and tested with the temperature solution fields obtained via Computational Fluid Dynamics (CFD) simulations. In the proposed methodology, the Kirchhoff-Helmholtz integral~\cite{godin1996} is implicitly evaluated in a convolutional layer and the CNN is trained considering different flow conditions at steady-state. For this reason, this framework can also be referred to as \textit{Physics-Informed Neural Networks} (PINNs), a term introduced by Raissi et al in Ref.~\cite{raissi2019}. Furthermore, the architecture of the CNN employed in the present study has been first presented in Ref.~\cite{ponciroli2021convolutional}.

The proposed model is developed to serve as a tool for real-time plant diagnostics which will ensure less invasive monitoring capabilities, thus improving nuclear reactor operation and reduction of maintenance costs for in-core sensors. The present approach is tailored for an implementation to advanced fast reactors, more specifically the Molten Salt Reactor (MSR) design. It should be noted that the direct coolant temperature measurement is not feasible for this type of reactor, mainly due to much higher operating values compared to the traditional light water reactors (LWRs). Hence, the proposed methodology features a promising technique. At the same time, given the generic nature of the algorithm, the proposed technique can be applied to the current nuclear fleet.

Different strategies of using Machine Learning (ML) techniques have been employed for characterizing physical systems, either by retrieving its unknown physical state based on limited and sparse information or by discovering the partial differential equations underlying them. Recently,  Raissi et al in Ref.~\cite{raissi2019} proposed accounting for the governing equations when evaluating the weights and bias of a deep learning model through an automatic differentiation scheme.  These authors claim that such approach helps minimize the residuals in the training stage rather than doing so without any prior information. This reference feature interesting studies as the resulting model should provide essential gains in the execution speed for computationally demanding problems.

%For instance, in Ref.~\cite{lagaris1998}, Lagaris et al proposed building a closed analytical solution for systems governed by either ordinary or partial differential equations through an approximation provided by feedforward neural networks.

%More recently,  Raissi et al in Ref.~\cite{raissi2019} proposed accounting for the governing equations when evaluating the weights and bias of a deep learning model through an automatic differentiation scheme, following what has been proposed in Ref.~\cite{baydin2015}}.  The authors of Ref.~\cite{raissi2019} claim that such approach helps to minimize the residuals when performing the training rather than doing so without any prior information, like Lagaris proposed in his earlier study. Both Refs.~\cite{lagaris1998, raissi2019} feature interesting studies as their resulting models should provide essential gains in the execution speed for computationally demanding problems.

%Recently,  Raissi et al in Ref.~\cite{raissi2019} proposed accounting for the governing equations when evaluating the weights and bias of a deep learning model through an automatic differentiation scheme.  These authors claim that such approach helps minimize the residuals in the training stage rather than doing so without any prior information, like Lagaris proposed in his earlier study. Both Refs.~\cite{lagaris1998, raissi2019} feature interesting studies as their resulting models should provide essential gains in the execution speed for computationally demanding problems.

Development has also been made on using ML for turbulence modeling.  Duraisamy \& Durbin in Ref.~\cite{durbin2014} employed a supervised Neural Network to model a bypass transition.  In this study, ML was used to model an intermittency function which is then blended in a closure Reynolds-Averaged Navier Stokes (RANS) formulation in order to account for a bypass process in turbulence transition. Just as a reference, the bypass process occurs such that the free-stream turbulence enters the boundary layer either through the leading edge or through interactions with the boundary layer from above~\cite{boiko1994}. In Ref.~\cite{durbin2014}, the authors focused on the second mechanism.

Refs.~\cite{jiang2021,wang2018} also used ML to model turbulence. Specifically in Ref.~\cite{jiang2021}, Jiang et al shift the attention from improving the conventional RANS closures and focused on building a new model by training a deep neural network with rich data from high-resolution simulations. Interestingly, the resulting model features a good generalization in both two and three-dimensional flows for different Reynolds numbers. 

%Ref.~\cite{raissi2019} feature interesting studies as their resulting models should provide essential gains in the execution speed for computationally demanding problems.

Additionally, in Ref.~\cite{kutz2013}, Kutz et al developed a framework based on both Compressive Sensing and ML techniques for characterizing and retrieving the pressure field of a flow passing around a cylinder with a limited pressure measurement. The strategy consisted of two tasks. First, the authors employed a Model Order Reduction (MOR) method named proper orthogonal decomposition (POD)~\cite{berkooz1993} in which the dynamics of the pressure field are projected from a higher-dimensional discretized system to a lower-dimensional one. Once obtained, the POD modes are organized in a library matrix built using a supervised ML algorithm. Then, the second task is performed in an online fashion, also referred to as field application. In the latter, a signal processing technique termed Compressive Sensing is employed in order to reconstruct the entire pressure based on the measurement on the surface of the cylinder.

% field given sparse measurements  based on the fact that the signal has a sparse representation on an appropriate basis~\cite{candes2006,tropp2007}.

With the exception of Ref.~\cite{kutz2013}, the hereinbefore mentioned references exploited ML techniques by addressing a wide range of physical problems such that the proposed models featured a better performance and with a similar robustness as the conventional approaches, e.g. the RANS closures. The focus of the present work is readily different from this type of application, here we focus on leveraging CNNs capabilities to reconstruct physical fields from collected sensor measurements.

As test cases, a simple but representative heated-channel driven by an incompressible fluid is used to verify the CNN capabilities.  Predictions using the CNN are made considering different scenarios for the flow conditions, this includes a series of tests at low (\(Pr\ll1.0\)), moderate (\(Pr=1.0\)) and high (\(Pr=10.7\)) Prandtl numbers. Besides that,  a demonstration case considering a MSR is also provided, showing that the CNN is able to handle even more complex geometries.

%The Computational Fluid Dynamics (CFD) code Nek5000~\cite{nek5000} was employed to obtain the temperature field solutions for the various conditions used in both training and testing stages. Nek5000 is an open-source code based on the spectral element method (SEM) introduced by Patera in Ref.~\cite{patera1984} and developed at Argonne National Laboratory (ANL). On the other hand, the CNN has been developed using Keras deep-learning framework~\cite{gulli2017}. Keras is also an open-source software library featuring a Python interface for artificial neural networks and runs on top of TensorFlow library~\cite{tensorflow2015}.

The present work is structured as follows: the CNN employed is described and discussed in the second section. Still in this section, the CFD simulations details are provided, including the flow conditions considered. Next, Section~\ref{sec:results} presents the predictions made for the test cases. Besides that, a sensitivity analysis is also carried out targeting to check the robustness of the CNN model in a real application scenario where noise is inevitably present. This analysis consisted of overlaying the test inputs with a normal distribution of random numbers. The idea of including these distributions is to mimic noise in the inputs at different levels, hence standard deviation corresponding to 5\% and 10\% of the original input values were considered. Finally, the conclusion section provides a summary of the current findings and future works proposal.

\section{Convolutional Neural Network}
\label{sec:cnn}

In Ref.~\cite{ponciroli2021convolutional} the authors proposed a Convolutional Neural Network targeting the reconstruction of scalar fields described by the Helmholtz equation over domains with homogeneous media and with no source.  Such equation describes the spatial distribution of the physical quantities, for instance, the temperature, which is characterized by diffusive and/or advective phenomena. 

Furthermore, one of the objectives of the CNN proposed is to address the task of monitoring a system treating it as a Boundary Value (BV) problem,  consisting of finding the solution of a given differential equation subjected to a set of Boundary Conditions (BCs) Ref.~\cite{russel1975}.  Eq.~\ref{eq:KH_integral} derives as an analytical solution to the Helmholtz equation.  Such solution assumes that \(u(\mathbf{x})\) in the field of interest and that we have the knowledge of both Dirichlet \(u(\mathbf{x'})\) and Neumann \(\partial u(\mathbf{x'})/ \partial \mathbf{n} \) boundary conditions on the boundary \(\Omega\) that enclosures the domain of interest.

\begin{equation}
  \label{eq:KH_integral}
  u (\mathbf{x}) = \oiint\limits_{\Omega} \left( u (\mathbf{x'}) \frac{\partial G(\mathbf{x},\mathbf{x'})}{\partial \mathbf{n}} 
   			- G(\mathbf{x},\mathbf{x'}) \frac{\partial u (\mathbf{x'})}{\partial \mathbf{n}} \right) d\Omega \quad \textrm{with} \; \mathbf{x'} \in \Omega
\end{equation}

In the above equation, \(G(\mathbf{x},\mathbf{x'})\) represents the Green's function, or Green's third identity, to the operator (\(L=\nabla^2 + k^2\)),  being \(\nabla^2\) the Laplacian operator and \(k\) the wavenumber value.  Finally, Eq.~\ref{eq:KH_integral} is referred to as the Kirchhoff-Helmholtz integral and as it will be discussed next,  it features a key aspect of the CNN employed in the present work.

\begin{figure}[htb]
%\begin{spacing}{1.0}
\centering
\includegraphics[scale=1.0]{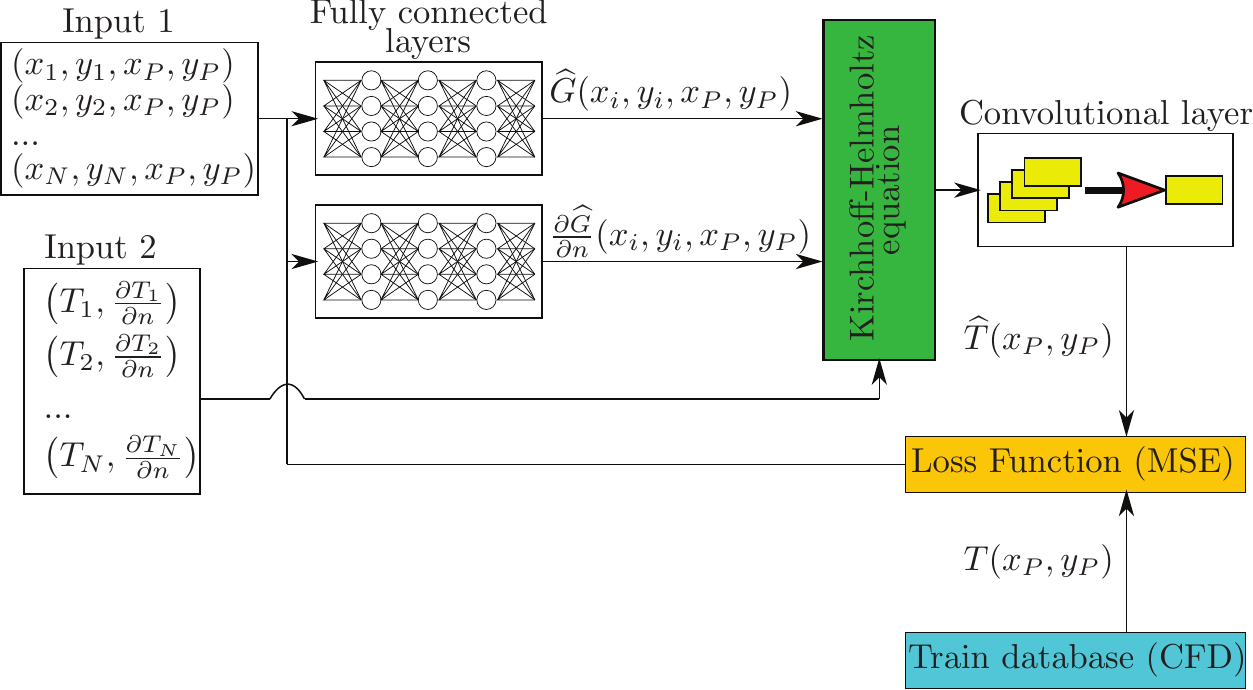}
\caption{Graphical representation of the Convolutional Neural Network.}
\label{fig:CNN_representation_2D}
%\end{spacing}
\end{figure}

In the present study, a similar CNN architecture from Ref.~\cite{ponciroli2021convolutional} is employed. This neural network is designed to reconstruct the spatial distribution of either two-dimensional or three-dimensional fields of physical quantities, here the temperature is considered.  The graphical representation of the network used is shown in Figure~\ref{fig:CNN_representation_2D}. As it can be observed from this figure,  a two-dimensional problem is considered for the present study as part of a simplification.  Furthermore, this model has been developed using Keras deep-learning framework~\cite{gulli2017}, an open-source software library featuring a Python interface for artificial neural networks.

% which runs on top of TensorFlow library~\cite{tensorflow2015}.

The proposed architecture receives two inputs, i.e. \textit{Inputs \#1} and \textit{\#2} shown in Figure~\ref{fig:CNN_representation_2D}. In \textit{Input \#1}, spatial features of the domain are provided by concatenating the coordinates \((x_i,y_i)\) and \((x_P,y_P)\), which respectively represent the collocation points where the boundary conditions are evaluated and an interior point where the field solution is known \textit{a priori}. Next, \textit{Input \#2} also results from concatenating data, this time with physical information rather than spatial. Hence, \textit{Input \#2} concatenates the Dirichlet \((T(x_i,y_i))\) and the Neumann \((\frac{\partial T}{\partial n}(x_i,y_i))\) boundary conditions in correspondence of the selected collocation points.

Similarly to what has been proposed in Ref.~\cite{ponciroli2021convolutional},  the \textit{a priori} unknown Green's function \(G\) is inferred through the supervised learning task of the proposed CNN. In this process, two sets of three fully connected layers each receives the data from \textit{Input \#1} in order to reconstruct the Green’s function and its normal derivatives, i.e. \(\widehat{G}(x_i,y_i, x_P, y_P)\) and \(\frac{\partial \widehat{G}}{\partial n}(x_i,y_i, x_P, y_P)\). Following, a physical layer shown in green allows calculating the integrand of the Kirchhoff-Helmholtz integral equation provided by Eq.~\ref{eq:KH_integral}.

Finally, a convolutional layer is used to integrate the output of the Helmholtz-Kirchhoff layer over the domain boundaries. Convolutional layers are traditionally placed at the beginning of CNNs to reduce images into a form that is easier to process, without losing important features which are critical for generating accurate predictions. In the current application, the convolutional layer constitutes the last step since it is used to map different contributions to the temperature field solution. The output of the convolutional layer is the estimated value of the field in correspondence to the interior points that are used for training/testing \(\widehat{T}(x_P,y_P)\). The learning process stems by minimizing the mean square error (MSE) between the predictions at the interior points \(\widehat{T}(x_P,y_P)\) and the actual solution value \(T(x_P,y_P)\).
%
%\vspace{4pt}
%\vspace{8pt}
%

Two points of this framework should be highlighted. Firstly, the present methodology is conceived to work under a supervised learning process. The goal of using this kind of learning is that the \textit{a priori} unknown Green's function values \(\widehat{G}\) and its normal derivatives \(\frac{\partial \widehat{G}}{\partial n}\) are assumed to be mapped when performing the backpropagation method~\cite{goodfellow2016} given the fact that the model accounts for the Kirchhoff-Helmholtz equation, Eq.~\ref{eq:KH_integral}. It should be noted however that the governing equations in the CFD calculation do not fully correspond to the Helmholtz equation once a convection term is also considered. Still, as it will be shown in the Results Section, the predictions made by the CNN which increases the influence of a diffusive process well suits as an approximation when compared to the CFD results.

%It should be noted however that this is not a valid assumption whenever the field solution does not respect a Helmholtz equation as previously stressed. This is a relevant constraint of the CNN regarding the cases being investigated in the present work, especially because the governing equations in the CFD calculation do not fully correspond to the Helmholtz equation once a convection term is considered in the former. 

%Interestingly,  the CNN features different performances depending on the flow conditions.  Better predictions are obtained for cases with lower Prandtl numbers (\(Pr\ll1.0\)) compared to moderate (\(Pr=1.0\)) and high Prandtl ones (\(Pr=10.7\)).  This will be shown in the Results Section and this feature is somehow expected given the fact that the low Prandtl numbers are diffusion dominated cases, differently from the other two cases. Thus, for (\(Pr\ll1.0\)) conditions the temperature 

The second point that should be stressed is the fact that there is no restriction on the choice for the batch size during the training process. This is a great advantage and is due to how the network is conceived. This feature can be understood by observing that the whole CNN model is meant to map both the spatial and physical data provided in \textit{Inputs \#1} and \textit{\#2} into a single temperature value at any part of the domain, i.e. \((x_P,y_P)\). In practical terms, this means that regardless the amount of data available for training, the batch size could be as low as the dimension of the prescribed solution being used when evaluating the loss function, i.e. \(T(x_P,y_P)\) which is simply a unity. Another interesting advantage about this feature is the fact that the proposed network can easily receive an active learning instead of a supervised learning. Differently from a pure supervised learning, where the training conditions are organized and supplied at the beginning of the training process, in active learning the algorithms interactively collect new training samples, which can be supplied to a pre-trained model.

Continuing, the Adaptive Moment Estimation (ADAM) method~\cite{kingma2015} was used as an optimization algorithm. Besides being computationally efficient once it features little memory requirements, ADAM well suits problems that are large in terms of data volume and number of parameters. Besides the optimization algorithm, the choice of activation function is also important to improve the accuracy of the predictions. Here the hyperbolic tangent function has been selected following what is suggested in Ref.~\cite{ponciroli2021convolutional}.

Additional features for training the models in the present study are the batch size of 32, the number of neurons per fully connected layer of 128, and 200 epochs in which the loss function is evaluated. These values were selected based on experiments and we do not claim them to be optimal but sufficient to retrieve the temperature field accurately. When tuning these parameters, special attention was given to avoid overfitting as it will be shown in the next section.

Finally, in order to improve the accuracy of the model, all the inputs data has been normalized to be within \([0,1]\) during the pre-processing stage. Differences in the scales across input variables, i.e. lengths and temperatures, may increase the difficulty of treating the problem such that large variations in the weight's values often lead to unstable models. Nevertheless, this issue can be avoided with a simple linear rescaling of the input variables as largely recommended in different references, e.g.~\cite{goodfellow2016,bishop1995}.

\subsection{Simulation details and flow conditions}
\label{sec:heated_channel}

\begin{figure}[!htb]
%\begin{spacing}{1.0}
\centering
\includegraphics[scale=0.6]{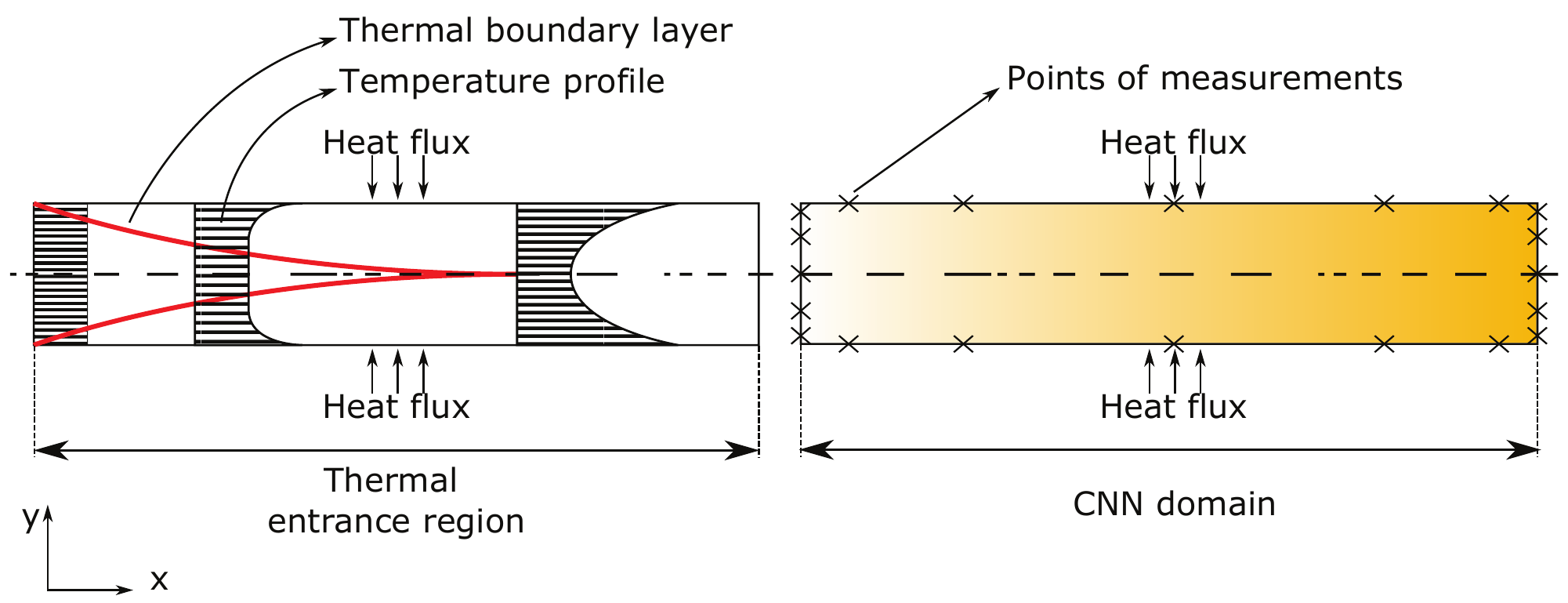}
\caption{Sketch of the proposed setup in which the CNN is tested.}
\label{fig:domain_scketch}
%\end{spacing}
\end{figure}

As a proof of concept, a simpler but representative heated channel driven by an incompressible fluid is considered in order to test the capabilities of the CNN instead of a more complex geometry of a nuclear reactor. The \(x\) variable represents the streamwise direction and the vertical direction is represented by \(y\). A sketch showing the problem setup is provided in Fig.~\ref{fig:domain_scketch}. Three scenarios considering different flow conditions for training and testing data were carried out in order to verify the CNN capabilities and robustness. In the first scenario the Reynolds is kept at a constant value of \(Re=10,000\) whereas the Prandtl changes, but only at low values, i.e. \(Pr\ll1.0\). In the second scenario, the Prandtl number is fixed at \(Pr\ll1.0\) while the Reynolds number changes, ranging from laminar to turbulent.  Finally, the third scenario also considers a fixed Prandtl, but now at a higher value of \(Pr=10.7\), hence resembling a molten salt flow condition. It should be noted that it is not affordable to train the model for all possible flow conditions. For this reason, a limited number of conditions are used to train the model as shown in Tab.~\ref{tab:conditions}. Still, the model should be able to make predictions for any conditions that is within the training set range, e.g. the test conditions from the same table.

As discussed in the previous section, the model employed should not account for non-linear effects in the temperature solution field, thus the region where the thermal boundary layer develops is used neither for training nor testing the network as shown in the sketch.
%It should be noted that the entrance region for the scenarios with \(Pr = 1.0\) and \(Pr = 10.7\) are much longer than for the case with \(Pr\ll1.0\) whereas the latter case features a much faster development of the thermal boundary layer when compared to the first two~\cite{ahtt5e}. Anyway, Tab.~\ref{tab:conditions} shows the conditions for the three scenarios studied.

%\vspace{8pt}
\begin{table}[!htb]
\centering
\caption{Flow conditions considered for train and test.}
\label{tab:conditions}
%\vspace{8pt}
\begin{tabular}{||c||c|c||c|c||c|c||} \hline \hline
\multicolumn{1}{||c||}{|} & \multicolumn{2}{|c||}{Low Prandtl (\(Pr\ll1.0\))} & \multicolumn{2}{|c||}{Moderate Prandtl (\(Pr=1.0\))}  & \multicolumn{2}{|c||}{High Prandtl (\(Pr=10.7\))} \\
\hline \hline
\multicolumn{1}{||c||}{Train sample} & \multicolumn{1}{|c|}{\ \(Re\)} & \multicolumn{1}{|c||}{\ \(Pr\)} & \multicolumn{1}{|c|}{\ \(Re\)} & \multicolumn{1}{|c||}{\ \(Pr\)} & \multicolumn{1}{|c|}{\ \(Re\)} & \multicolumn{1}{|c||}{\ \(Pr\)}
 \\ \hline \hline
 1 &   10,000 &  6.6E-04  & 100 & 1.0  & 10,000 & 10.7  \\ \hline
 2 & 10,000  & 1.0E-03 & 200  & 1.0  & 12,500 & 10.7  \\ \hline
 3 & 10,000  &  2.0E-3 & 400  & 1.0  & 15,000 & 10.7  \\ \hline
 4 & -  & - & 800 & 1.0  & 17,500 & 10.7  \\ \hline
 5 & -  & - & 1,600 & 1.0  & 20,000 & 10.7  \\ \hline
 6 & -  & - & 3,200 & 1.0  & - & -  \\ \hline
 7 & - & - & 6,400 & 1.0  & - & -  \\ \hline
 8 & - & - & 12,800  & 1.0  & - & -  
\\ \hline \hline
\multicolumn{1}{||c||}{Test sample} & \multicolumn{1}{|c|}{\ \(Re\)} & \multicolumn{1}{|c||}{\ \(Pr\)} & \multicolumn{1}{|c|}{\ \(Re\)} & \multicolumn{1}{|c||}{\ \(Pr\)} & \multicolumn{1}{|c|}{\ \(Re\)} & \multicolumn{1}{|c||}{\ \(Pr\)}
 \\ \hline \hline
 1 &  10,000  & 8.0E-04 & 1,000 & 1,0  & 10,500 & 10.7 \\ \hline
 2 &  10,000  & 1.3E-03 & 2,000 & 1,0  & 14,000 & 10.7 \\ \hline
 3 &  -  &  - & 5,000 & 1,0  & 18,000 & 10.7  \\ \hline
 4 &  - & - & 10,000 & 1,0  & - & -  
\\ \hline \hline
\end{tabular}
\end{table}
%\vspace{8pt}

All data used in the present study was obtained via CFD simulation of incompressible flow using Nek5000~\cite{nek5000v19}, an open-source code based on the spectral element method (SEM) introduced by Patera in Ref.~\cite{patera1984} and developed at Argonne National Laboratory (ANL). In these simulations, the \(k-\omega\)~\cite{wilcox1993} RANS model has been employed to account for turbulence at sufficiently high Reynolds numbers. Nek5000 has received extensive validation in numerous references, for instance Refs.~\cite{merzari2013,shaver2020}. The considered channel has been heated with a prescribed flux on both wall boundaries as shown in Fig.~\ref{fig:domain_scketch}. Furthermore, the \(y^+<1.0\) criteria have been used when designing the mesh to ensure quality results. As a reference, the authors of Ref.~\cite{ponciroli2021convolutional}, in which the CNN has been first proposed, carried out their study using the Boundary Element Method (BEM)~\cite{beer2018} to obtain the solution fields employed as training data. In that reference, the study was carried out considering problems governed by the Laplace and the
Helmholtz equations.

As boundary conditions, a unitary heat flux on both walls has been considered for all the cases. On the inlets, a parabolic velocity profile with a mean value of \(\bar{u}=1\) and a uniform temperature of zero were employed.

It should be noted that the sketch shown in Fig.~\ref{fig:domain_scketch} is just a graphical representation while the number and positions of the points where the measurements are taken are just illustrative. The actual collocation points used to make the measurements were obtained using the Gauss-Lobatto-Legendre quadrature weights of \nth{6} order over the domain boundaries, meaning that in total 24 points were used. Furthermore, it should be stressed that the CNN model has been built using all the grid points from the CFD mesh. In other words, all the grid points of the mesh and their respective temperature solutions were supplied to the CNN when performing the supervised learning. This way, more than 20,000 grid points have been mapped by the model, hence providing the same spatial resolution as the original train data.

%
%\vspace{8pt}
%\vspace{8pt}
%

\subsubsection{MSR demonstration case}
\label{sec:demo_msr}

Besides testing the CNN at different conditions for a simple heated channel setup as proposed in Section~\ref{sec:heated_channel}, the CNN has been also employed to reconstruct the temperature field considering a more complex geometry of a MSR. The specific reactor model considered for the present study is based on concept proposed under the Euratom EVOL project~\cite{yamaji2014,rouch2014}. 

MSRs features a promising technology once it offers major advantages in safety and efficiency compared to the current fleet of Light Water Reactors (LWR). Both of these designs control fission to produce steam that powers electricity-generating turbines. However, there are some differences in how MSRs are conceived. The main difference between these reactors is the fact that molten salt is used as a coolant in MSRs rather than water such as in LWRs.  Besides that, the former has the nuclear fuel dissolved in the coolant instead of having it displaced in fuel rods. These features provide benefits including significantly enhanced efficiency, load following, and the ability to operate at high temperatures, which also makes them suitable for non-electric applications but where high heat input is required.

\begin{figure}[!htb]
%\begin{spacing}{1.0}
\centering
\hspace*{-2cm} 
\includegraphics[scale=0.46,valign=t]{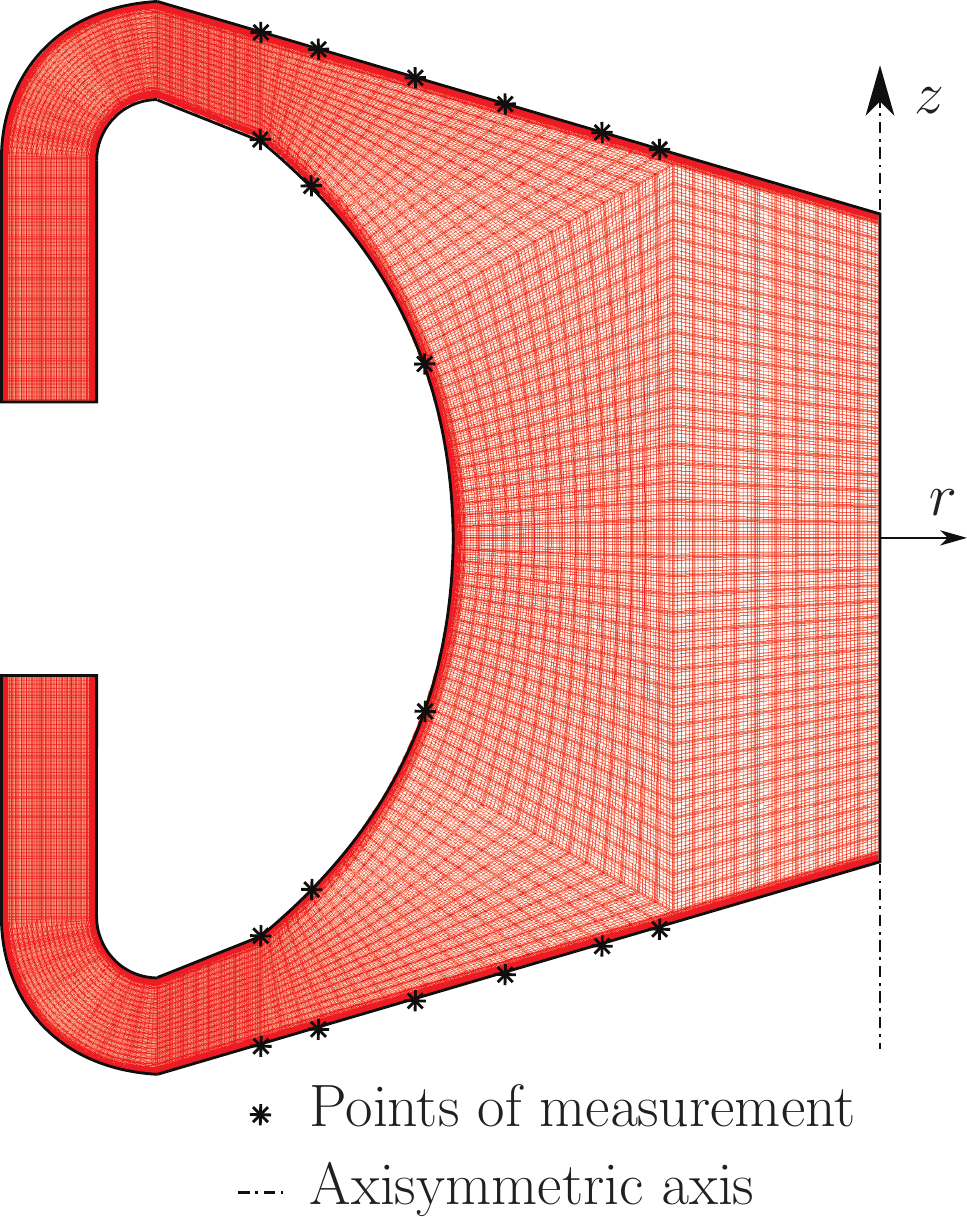}
\hspace*{2cm} 
\includegraphics[scale=0.46,valign=t]{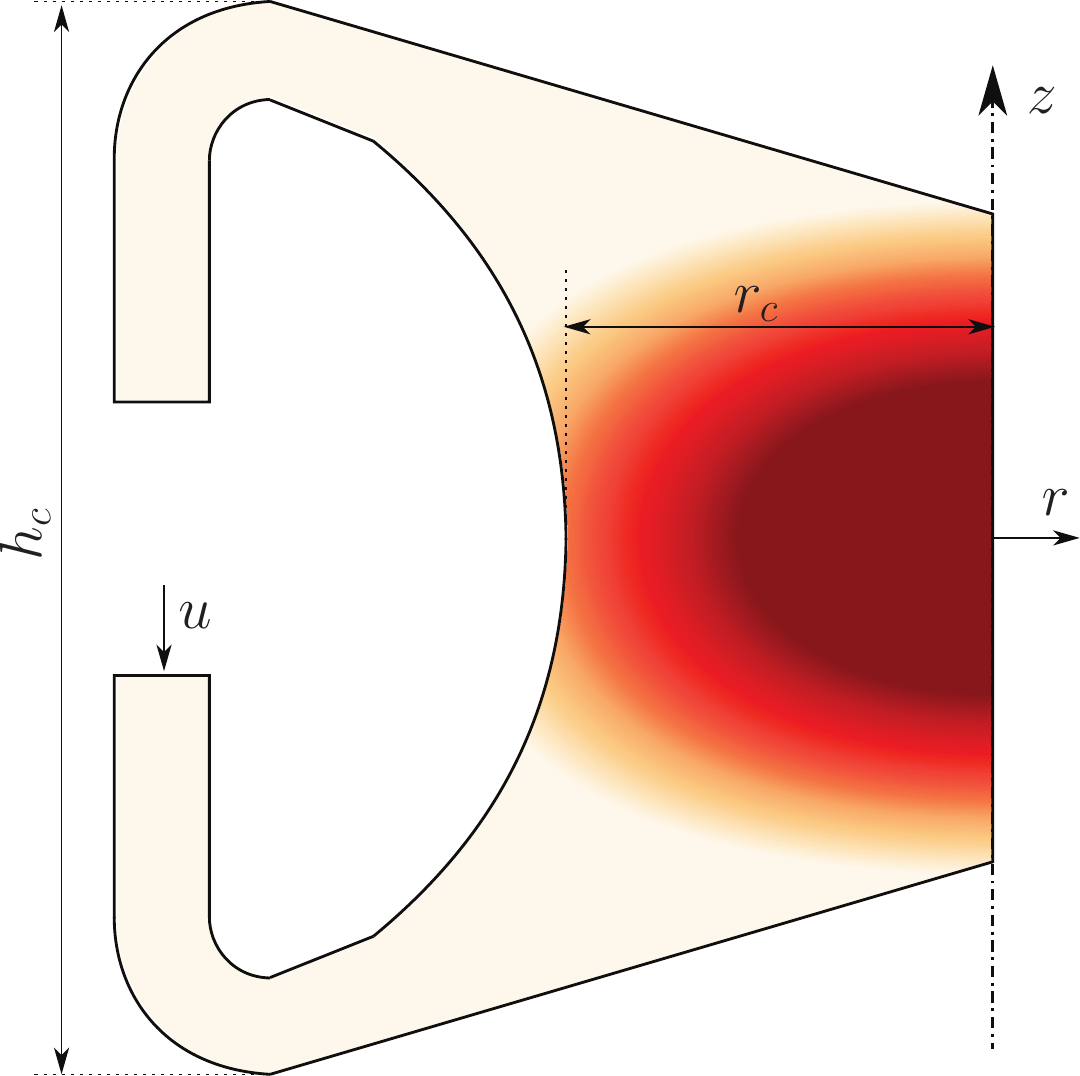}
\caption{Application of the proposed CNN for a more complex case of a MSR: on the left, the CFD mesh is shown along the measurement points, on the right: the power profile employed is presented.}
\label{fig:msr_cnn}
%\end{spacing}
\end{figure}

Fig.~\ref{fig:msr_cnn} shows: (a) the CFD mesh and the sensors used for temperature reconstruction (b) power profile considered following a cosine distribution in both radial and axial directions.  As it can be seen, the CFD model features a two-dimensional setup with respect to an axisymmetric axis (\(z\)) of a MSR core cavity.  In Fig.~\ref{fig:msr_cnn} (a) we notice that a total of 18 points of measurements are selected (\(*\) marks). Furthermore, special attention is given close to the hot spot region, i.e. the center of the top core-cavity wall, where the maximum temperature is expected to occur. Measurement points in this region should be avoided due to the high temperatures that the fuel may exhibit, thus the CNN should not depend on measurements in this location. Still, the CNN should have the capability to satisfactory predict the peak temperature in that region based on the set of measurement points remotely arranged.

%
%\vspace{8pt}
%\vspace{8pt}
%

The MSR CFD model employed in the present work features the simplified two-dimensional version of the Large Eddy Simulation (LES) studied in Ref.~\cite{leite2021}. The core has a height of \(h_c=1.6 \; m\) along the centerline and width of \(2.65 \; m \) in the transverse direction. The minimum core-cavity radius is \(r_c=1.05 \; m\) and it increases up to \(1.53 \; m\). The geometry is reminiscent of an hourglass shape such that the peripheral wall is a curved surface.  This aspect is essential in this core concept once it ensures a relatively uniform velocity distribution inside the cavity.

The mesh shown in Fig.~\ref{fig:msr_cnn} (b) has 3,700 second-order quadrilateral elements. Moreover, this mesh is represented as a tensor-product of \(7^{th}\) order Lagrange polynomials built on $8$ Gauss Lobatto Legendre collocation points in each direction.  The same mesh was employed in Ref.~\cite{leite2021} in order to ensure sufficient spatial resolution.

Similar to the heated channel setup proposed in Section~\ref{sec:heated_channel}, turbulence is modeled using the \(k-\tau\) RANS closure. Training and predictions are performed for a single flow condition with the purpose of demonstrating the feasibility of using the proposed CNN as a diagnostic tool in a more complex geometry of a MSR. The Prandtl number considered is \(Pr=10.7\), featuring a high value typically seen in MSRs.  Furthermore,  the corresponding Reynolds number is around \(18,400\), which is defined based on the mean velocity through the minimum core diameter, i.e. \(D_c=2 r_c\), Fig.~\ref{fig:msr_cnn}. Finally,  a dimensionless temperature field is evaluated resulting from a heat source following a cosine function provided by Eq.~\ref{eq:power_profile}:

\begin{equation}
  \label{eq:power_profile}
  q'''(r,z) = q'''_{max}cos\left(\frac{\pi}{h_c}z\right) \cdot cos\left(\frac{\pi}{r_c} r\right)
\end{equation}

Where the heat source peak is simply \(q'''_{max}=1.0\) as a part of a normalization.

\section{Results and Discussion}
\label{sec:results}

In the present section, the predictions made for the test cases considering the three scenarios from Tab.~\ref{tab:conditions} are reported and discussed.  Besides that, a sensitivity analysis is also carried out in order to check the robustness of the CNN. Such analysis consisted of overlaying the original test inputs with a normal distribution of random values in which the standard deviation corresponds to \(\sigma=5\%\) and \(\sigma10\%\) of the original input values. Moreover, the results for the MSR demonstration use of the CNN is also presented.

\begin{figure}[!htb]
%\begin{spacing}{1.0}
\centering
\includegraphics[scale=0.5]{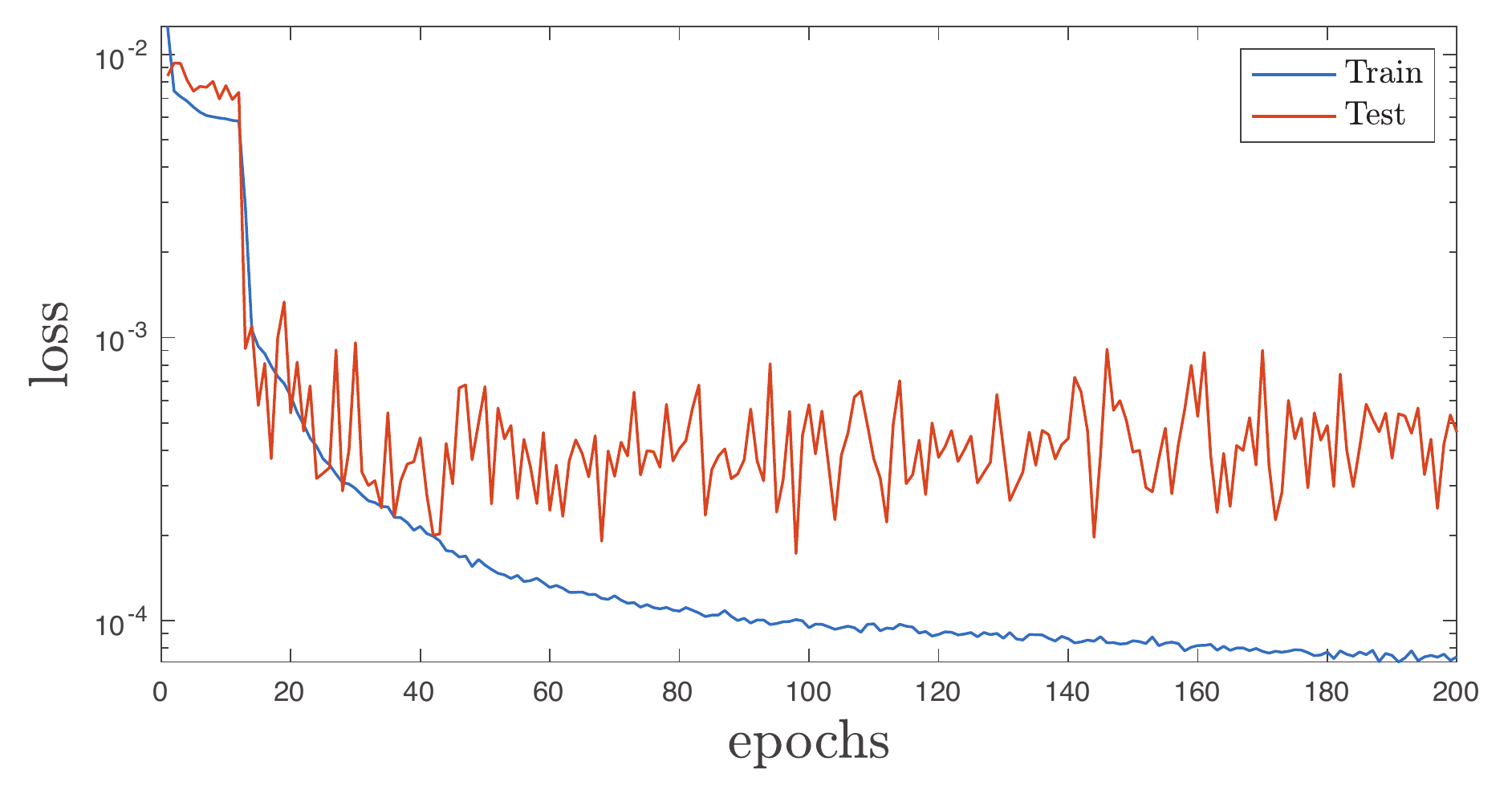}
\caption{The loss function for the supervised learning considering \(Pr=1.0\) scenario.}
\label{fig:loss}
%\end{spacing}
\end{figure}

Furthermore, Fig.~\ref{fig:loss} shows the loss function throughout the epoch numbers obtained in the supervised learning for the \(Pr=1.0\) expedition. The same plot considering for the other cases are not reported for the sake of brevity, but they feature the same behavior from this figure. The training error is shown in blue and the testing error in red, both as a function of the 
epoch numbers. In supervised learning algorithms, a good indication that overfitting may have occurred is when the test error increases whereas the training error still decrease or remains still. However, this is not the trend observed, thus the trained model should not suffer from overfitting. At the same time, the test loss converged to a minimum value, meaning that the model meets an optimum solution.

%
%\vspace{8pt}
%\vspace{8pt}
%

\subsection{Low Prandtl results, \(\mathbf{Pr\ll1.0}\)}
\label{sec:low_Pr}

\begin{figure}[htb]
    \centering % <-- added
\begin{subfigure}{0.5\textwidth}
  \hspace{0.4cm}
  \includegraphics[scale=0.45,trim={110 200 140 220},clip]{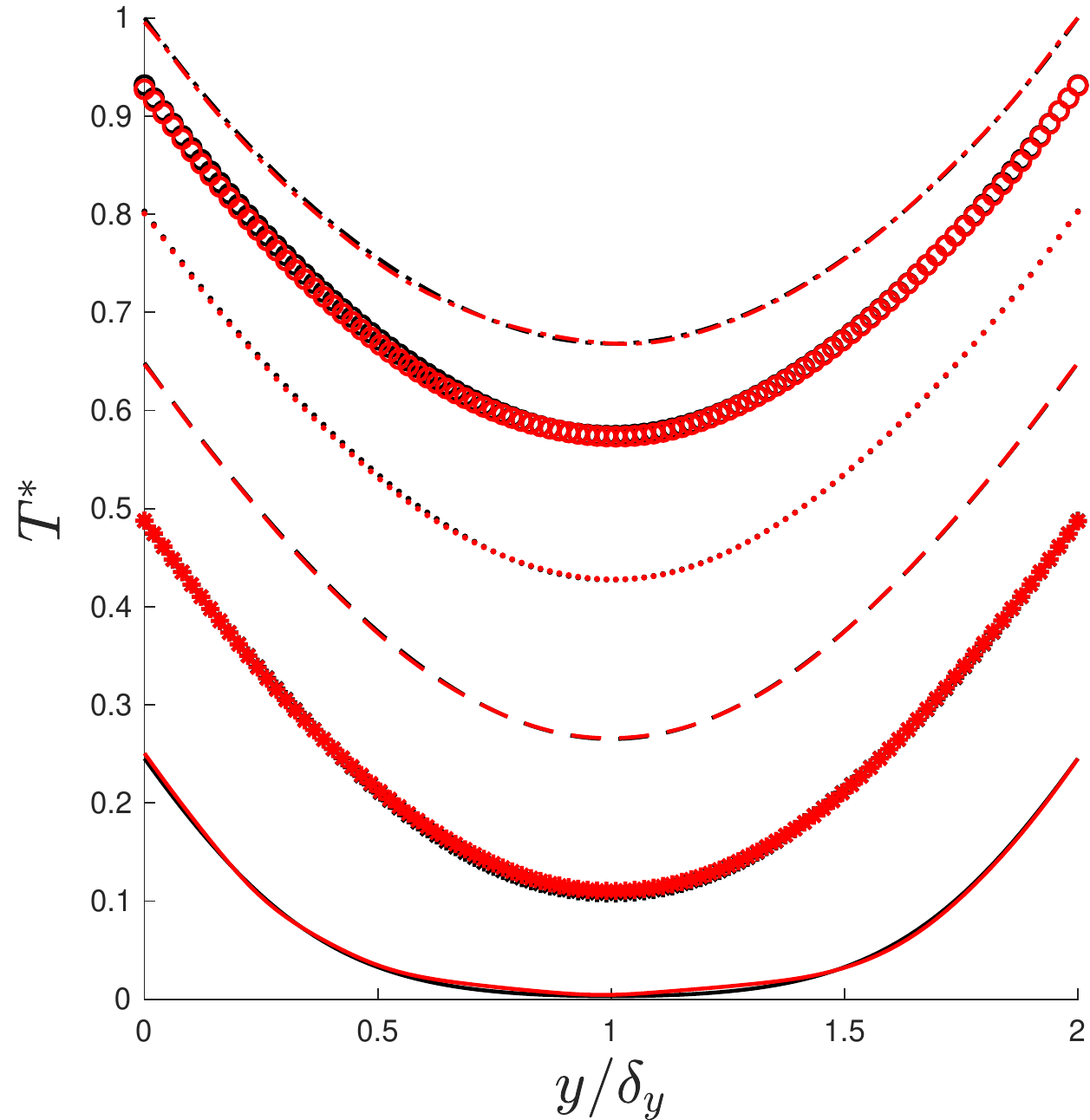}
  \caption{}
  \label{fig:T_lineplots_lowPr_case_1}
\end{subfigure}\hfil % <-- added
\begin{subfigure}{0.5\textwidth}
  \hspace{0.4cm}
  \includegraphics[scale=0.45,trim={110 200 140 220},clip]{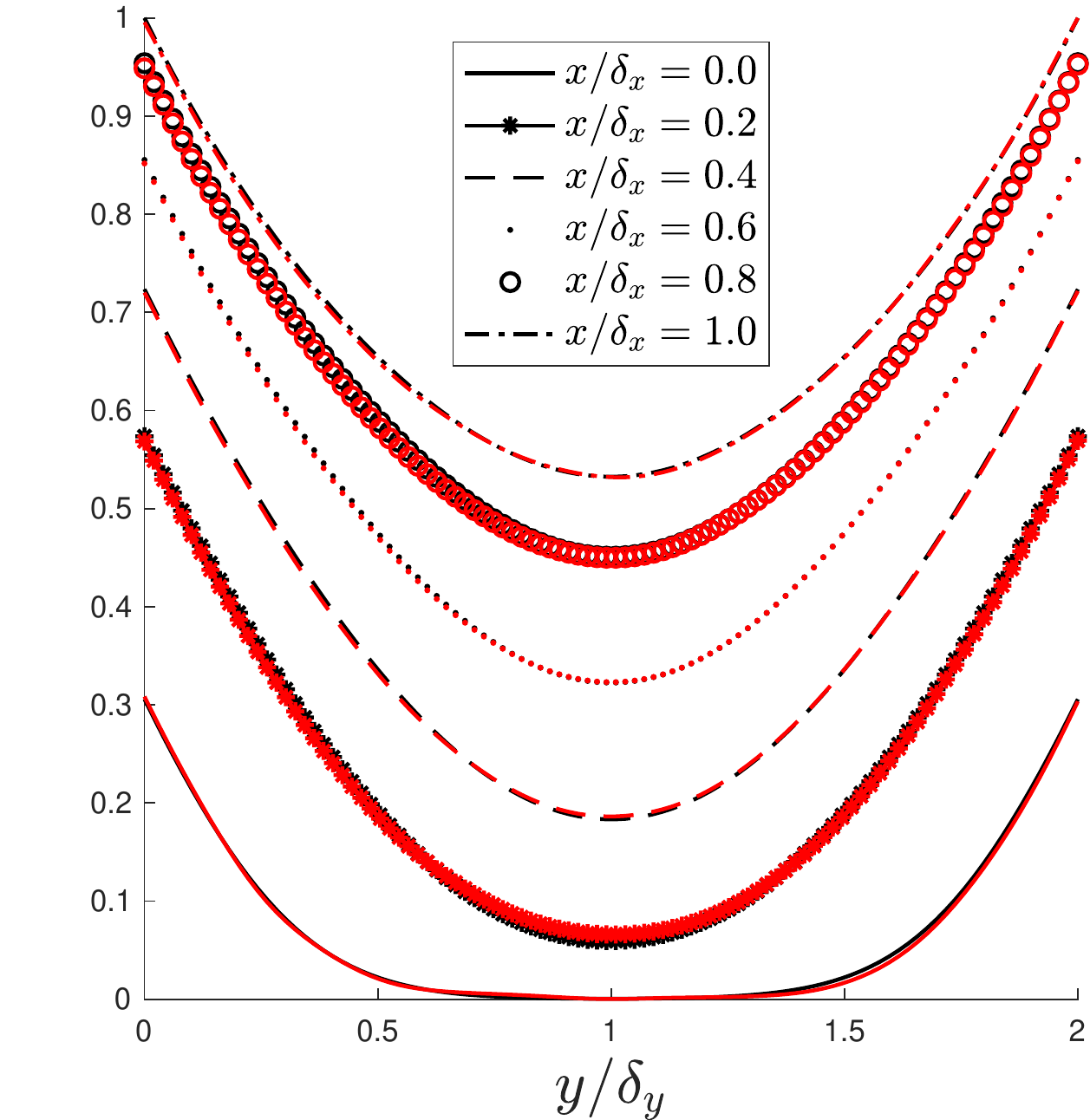}
  \caption{}
  \label{fig:T_lineplots_lowPr_case_2}
\end{subfigure}\hfil % <-- added
\caption{Temperature line plots for test cases with \(Pr\ll1.0\) along different positions of the heated channel: (a) test case 01 and (b) test case 02. The red lines represents the predictions and the black lines represents the CFD solutions.}
\label{fig:T_lineplots_lowPr}
\end{figure}

The temperature profiles at different streamwise positions of the channel are shown in Fig.~\ref{fig:T_lineplots_lowPr} for the two test cases from \(Pr\ll1.0\) scenario, Tab~\ref{tab:conditions}. The red lines show the predictions made by the CNN while the black lines represent the solutions obtained via CFD simulations using Nek5000.  Both spatial variables \(x\) and \(y\) are normalized respectively by the length of the channel \(\delta_x\) and by the half-height of the channel \(\delta_y\) and the temperatures \(T^*\) are given in dimensionless units henceforth, Eq.~\ref{eq:T_dimensionless}.

\begin{equation}
  \label{eq:T_dimensionless}
  T^*=\frac{T(x,y)-T_{min}}{T_{max}-T_{min}}
\end{equation}

Interestingly, the CNN is capable of retrieving an almost identical temperature field as the solution provided by Nek5000.

\subsubsection{Sensitivity analysis with 5\% and 10\% levels of randomness for \(\mathbf{Pr\ll1.0}\)}
\label{sec:sensitivity_modPr}

As part of a sensitivity to test the robustness of the CNN, the same inputs used to obtain the results shown in Fig.~\ref{fig:T_lineplots_lowPr} were employed, however, this time they are overlaid with random values following a normal distribution in which the standard deviation corresponds to \(\sigma=5\%\) or \(\sigma=10\%\) of the original inputs. The goal of this analysis is to quantitatively evaluate how the CNN could perform in a real application, where noise is inevitably present in the measurement points.

\begin{figure}[htb]
    \centering % <-- added
\begin{subfigure}{0.5\textwidth}
  \includegraphics[scale=0.35,trim={20 25 30 50},clip]{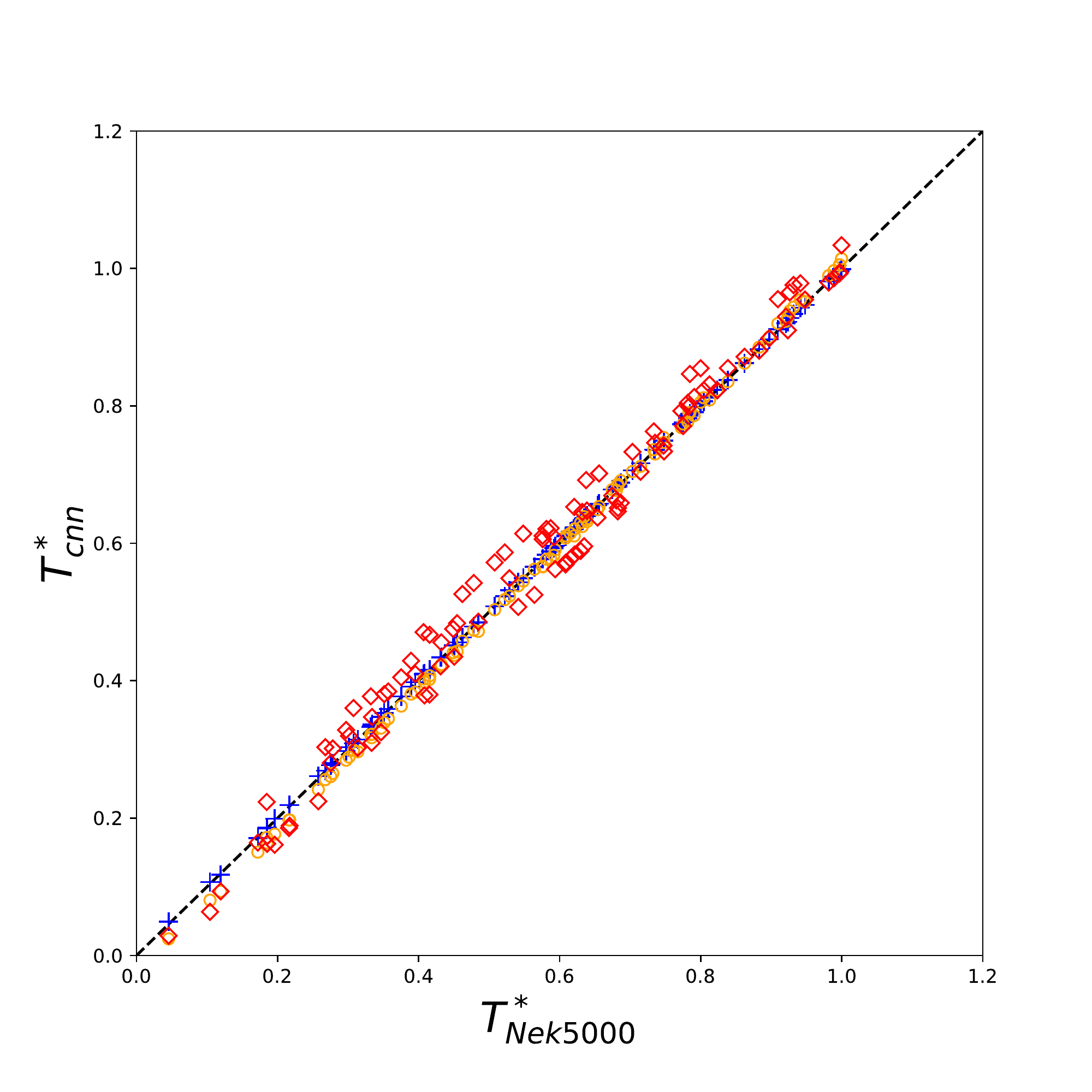}
  \caption{}
  \label{fig:noise_lowPr_01}
\end{subfigure}\hfil % <-- added
\begin{subfigure}{0.5\textwidth}
  \includegraphics[scale=0.35,trim={20 25 30 50},clip]{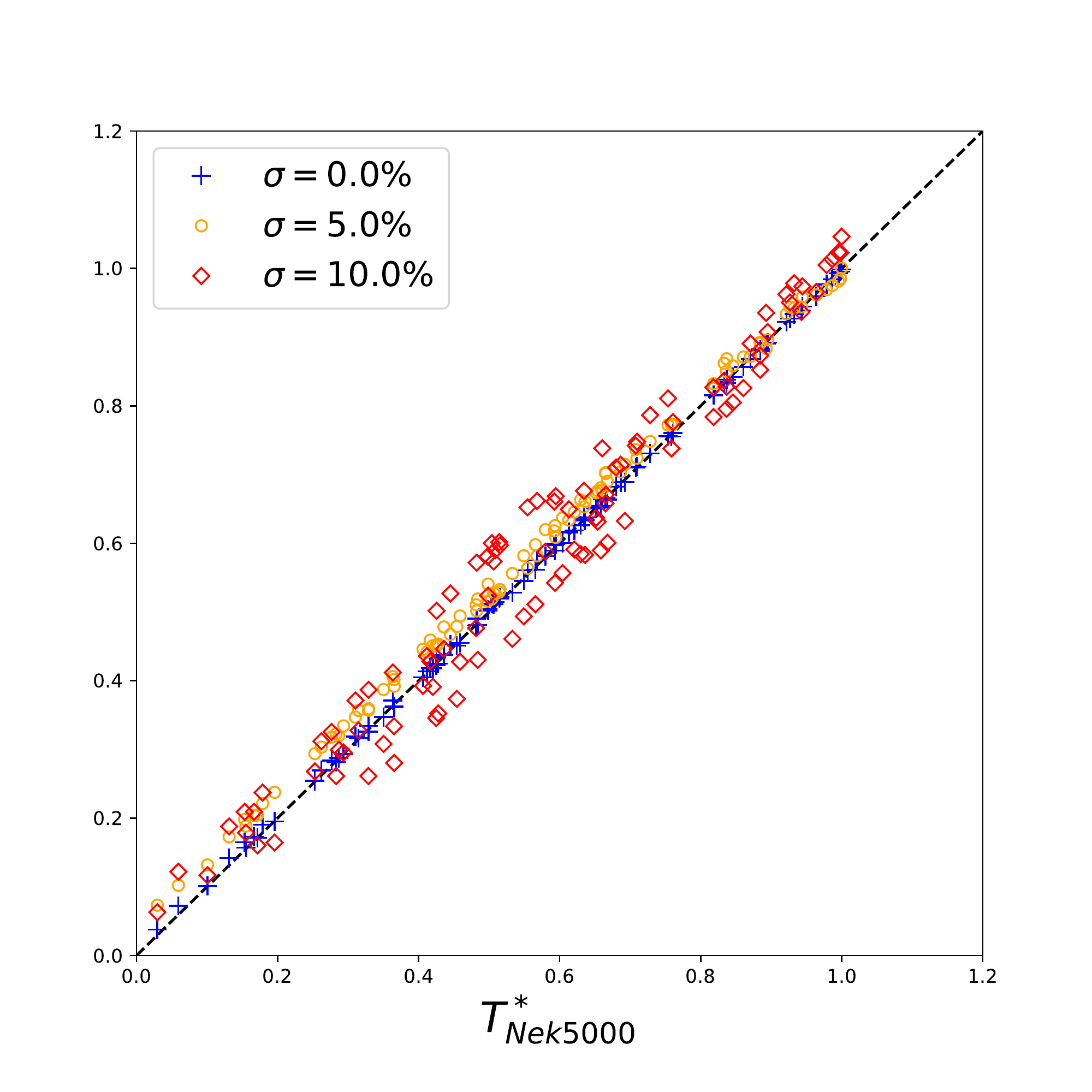}
  \caption{}
  \label{fig:noise_lowPr_02}
\end{subfigure}\hfil % <-- added
\caption{Predictions made by the CNN at low Prandtl numbers (\(Pr\ll1.0\)): (a) test case 01 and (b) test case 02.  The plots show results considering different values for \(\sigma\) as standard deviation at random locations within the domain.}
\label{fig:noise_lowPr}
\end{figure}

The plots shown in Fig.~\ref{fig:noise_lowPr} are provided to aid in visualizing the impact of including randomly distributed values into the test inputs. These figures show the scatter plots of a limited number of points within the domain, e.g.  50, showing the predictions made by the CNN versus the actual solution obtained using Nek5000 for test cases 01 and 02, Tab.~\ref{tab:conditions}.  Ideally, all the predicted values should be as close as possible to a regressed diagonal which is shown in black lines throughout this study.

\subsection{Moderate (\(\mathbf{Pr = 1}\)) and high (\(\mathbf{Pr = 10.7}\)) Prandtl results}

\begin{figure}[htb]
    \centering % <-- added
\begin{subfigure}{0.5\textwidth}
  \includegraphics[scale=0.45,trim={110 200 130 235},clip]{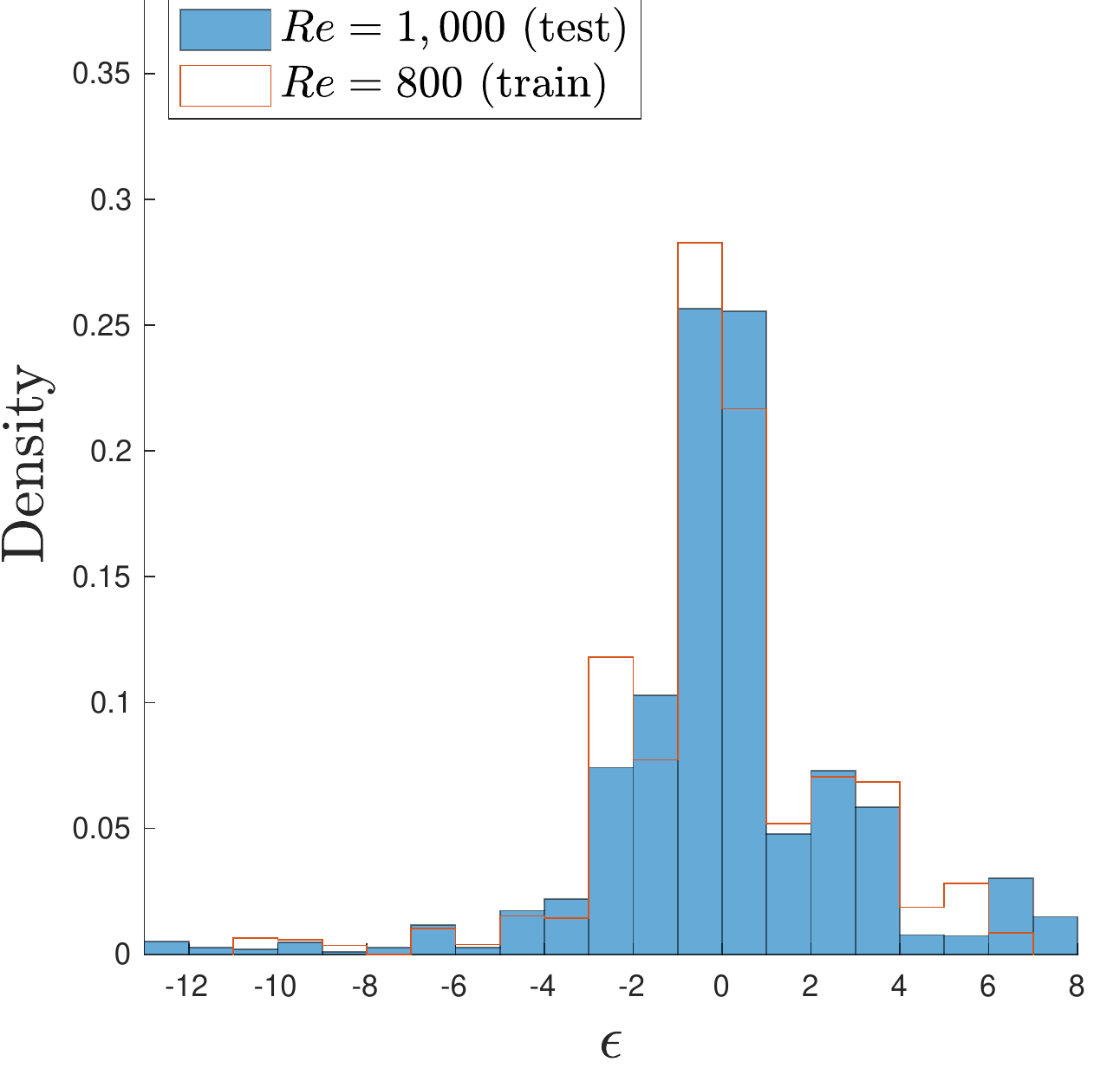}
  \caption{}
  \label{fig:histogram_modPr_01}
\end{subfigure}\hfil % <-- added
\begin{subfigure}{0.5\textwidth}
  \includegraphics[scale=0.45,trim={110 200 130 235},clip]{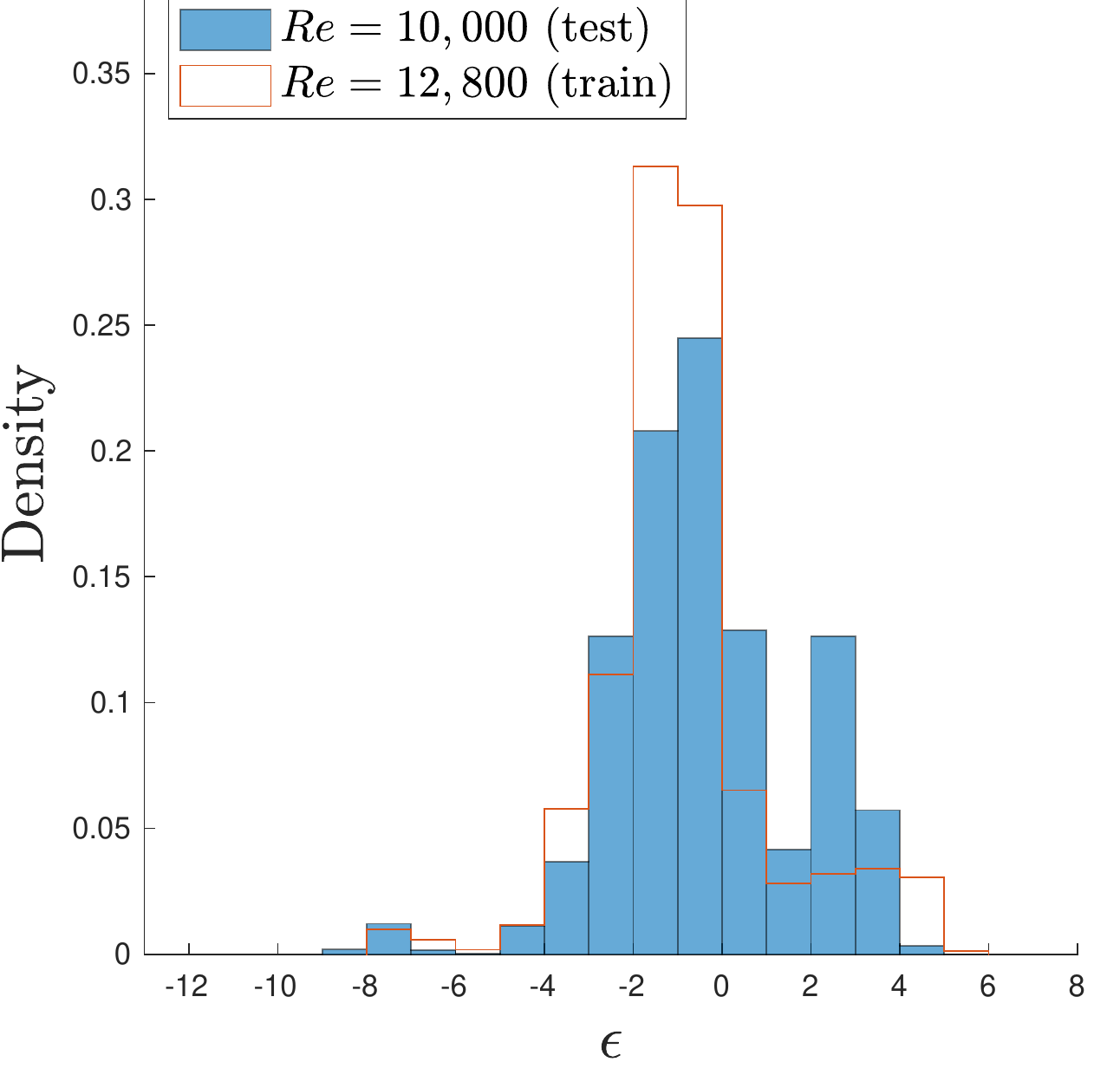}
  \caption{}
  \label{fig:histogram_modPr_04}
\end{subfigure}\hfil % <-- added
\caption{Histogram of absolute errors for cases with \(Pr=1.0\): (a) test case 01 (\(Re=1,000\)) overlaid with train case 04 (\(Re=800\)) and (b) test case 04 (\(Re=10,000\)) overlaid with train case 08 (\(Re=12,800\)).}
\label{fig:histogram_modPr}
\end{figure}

Differently from the temperature line plots presented in Section~\ref{sec:low_Pr}, Fig.~\ref{fig:histogram_modPr} shows the histogram of the signed errors considering the highest and the lowest Reynolds numbers tested in \(Pr=1.0\) scenario. The test cases considered in these histograms are respectively at \(Re=1,000\) and \(Re=10,000\) flow conditions, Tab.~\ref{tab:conditions}. Besides that, each of these plots are overlaid with the errors distributions from their closest train conditions. In that case, the flows are respectively at \(Re=800\) and \(Re=12,800\), also from Tab.~\ref{tab:conditions}. 

In general, the density in these histograms are centered at \(\epsilon=0\) in all cases considered,  also including the high Prandtl number cases \(Pr=10.7\) altough they are not being shown for brevety. Moreover, the poorer performance of the CNN for the test cases compared to the training ones is an expected result and it is reasonable to verify this fact.

It should be noted that the moderate Prandtl number cases features the particularity of accounting for both laminar and turbulent flow conditions as the Reynolds ranges from \(Re=100\) up to \(Re=12,800\), see Tab.~\ref{tab:conditions}.  Interestingly, despite the challenge of modeling such non-linear behavior, the CNN proved to be capable of retrieving these conditions with a good level of accuracy. 

%DEVE SE NOTAR AINDA QUE EXISTE UMA PARTICULARIDADE PARA O GRAFICO APRESENTADO EM (B) AONDE A CNN ACABA FAZENDO PREDICOES PIORES PARA A CONDICAO DE TESTE COMPARATIVAMENTE AO MOSTRADO EM (A).  ESSA DISPARIDADE PROVAVELMENTE SE DEVE AO FATO DAS CONDICOES DE TREINAMENTO ESTAREM MUITO ESPACADAS PARA CASOS MAIS TURBULENTOS. POR ISSO ESSA DEVE SER UMA CARACTERISTICA NAO DA CNN EM SI, MAS DAS CONDICOES CONSIDERADAS PARA O TREINAMENTO DELA.

\subsubsection{Sensitivity analysis with 5\% and 10\% levels of randomness for moderate and high Prandtl numbers}

\begin{figure}[htb]
    \centering % <-- added
\begin{subfigure}{0.5\textwidth}
  \includegraphics[scale=0.35,trim={20 25 30 50},clip]{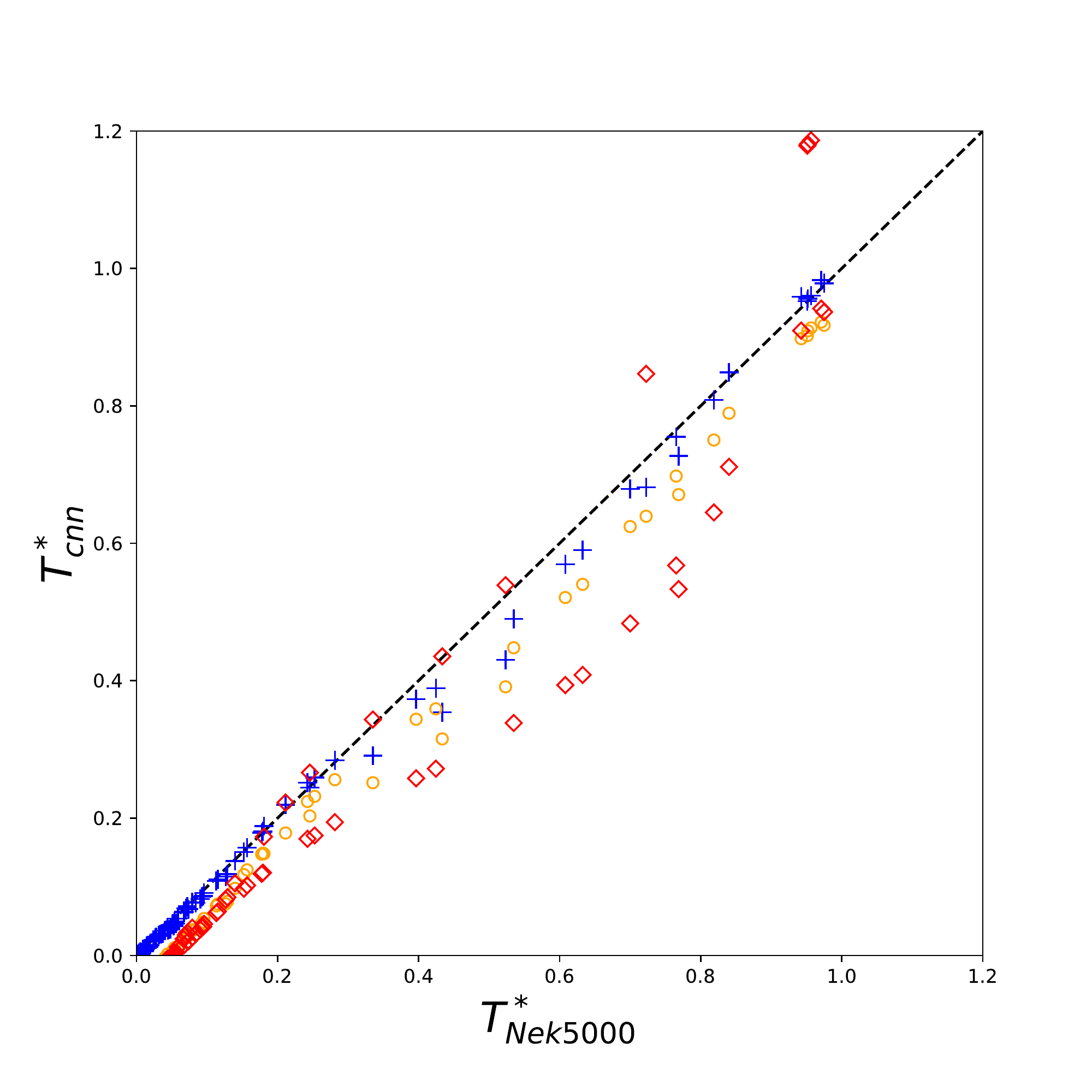}
  \caption{}
  \label{fig:noise_highPr_01}
\end{subfigure}\hfil % <-- added
\begin{subfigure}{0.5\textwidth}
  \includegraphics[scale=0.35,trim={20 25 30 50},clip]{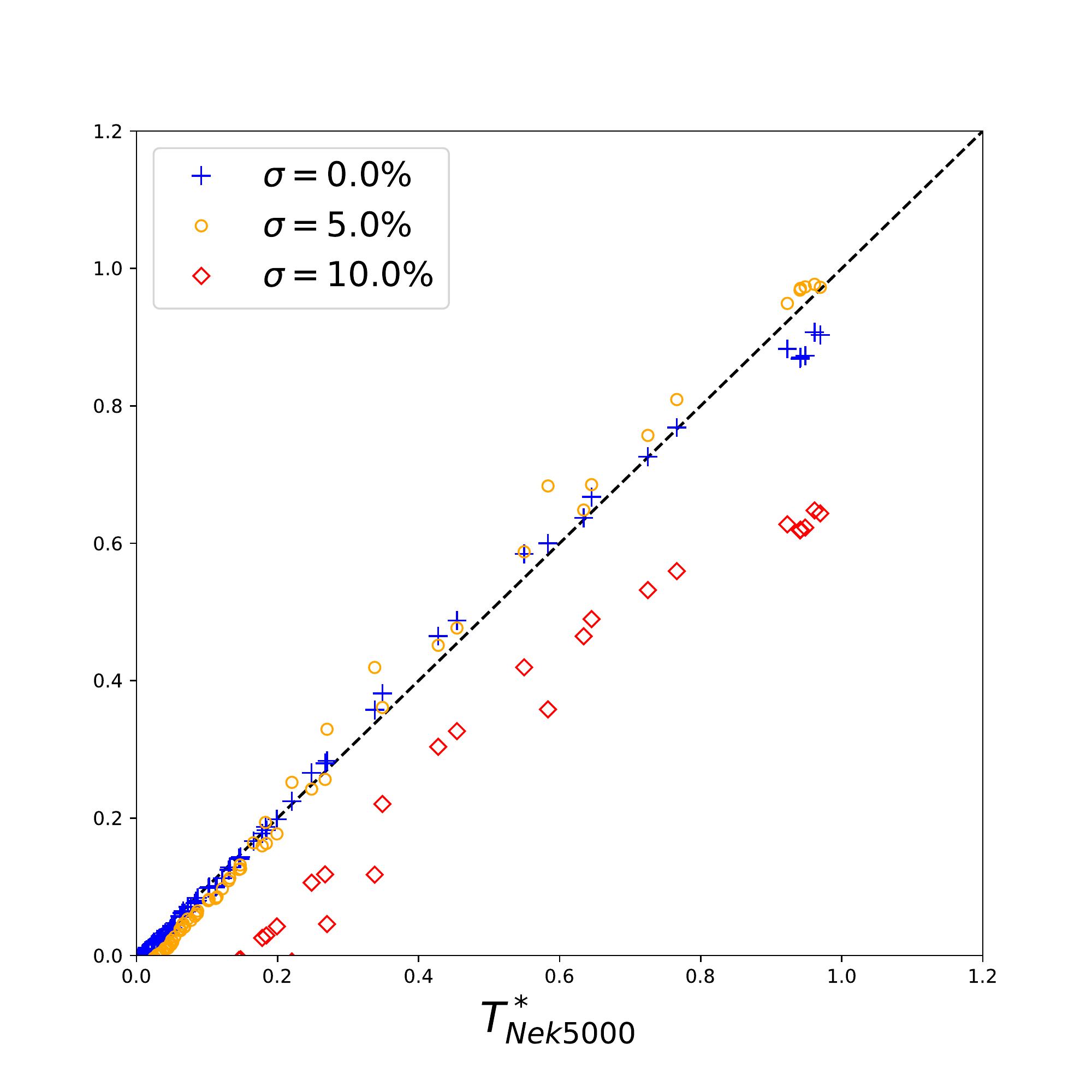}
  \caption{}
  \label{fig:noise_highPr_03}
\end{subfigure}\hfil % <-- added
\caption{Predictions made by the CNN at high Prandtl numbers (\(Pr=10.7\)): (a) test case 01 and (b) test case 03.  The plots show results considering different values for \(\sigma\) as standard deviation at random locations within the domain.}
\label{fig:noise_highPr}
\end{figure}

A similar sensitivity analysis as the one described in Section~\ref{sec:sensitivity_modPr} is also carried out for both moderate and high Prandtl numbers scenarios. Similarly to what has been discussed in that section, Figs.~\ref{fig:noise_highPr} presents the impact of including \(\sigma=5\%\) and \(\sigma=10\%\) of the test input values as the standard deviation of a random distribution. For the sake of brevity, only the plots considering the test cases 01 and 03 at \(Pr=10.7\) are reported here, but it should be noted that all the other cases performed in a similar manner.

%
%\begin{figure}[!htb]
%\centering
%\hspace*{-1cm} 
%\includegraphics[scale=0.43,trim={20 25 30 50},clip]{figures/noise_modPr_2}
%\hspace*{0cm} 
%\includegraphics[scale=0.43,trim={20 25 30 50},clip]{figures/noise_modPr_3}
%\caption{Predictions made by the CNN for test cases 02 and 03 at moderate Prandtl number (\(Pr=1.0\)) versus the temperature values from the CFD solution. The plots show results considering different values for \(\sigma\) as standard deviation at random locations within the domain.}
%\label{fig:noise_modPr}
%\end{figure}
%

%\subsection{Results summary}
%\label{subsec:summary}

When comparing the plots from Fig.~\ref{fig:noise_lowPr} and~\ref{fig:noise_highPr} it is clear to observe better predictions made by the CNN in low Prandtl numbers cases.  In this kind of regime, the molecular diffusivity of heat is higher than the molecular diffusivity of momentum whereas in high Prandtl conditions the opposite is observed.  Because of that, the temperature solution field for cases exhibiting \(Pr\ll1.0\) is somehow closer to a solution provided by a pure diffusive process, which in turn can be expressed in the form of the Helmholtz equation. With that, it is remarkable that the studied CNN favors flow regimes that are closer to systems that are better characterized by diffusion.

Additionally, the performance of the CNN is also measured in terms of the mean (\(\overline{L_2}\)) and the maximum (\(L_\infty\)) Euclidian norms stemmed from the difference between the actual solution and the predictions made by the CNN. Eqs.~\ref{eq:l2norm} and~\ref{eq:linfnorm} provides both quantities.

\begin{equation}
  \label{eq:l2norm}
  \overline{L_2}=\sqrt{\sum_{i=1}^{N}\frac{|T_{cnn}(x_i,y_i)-T_{Nek5000}(x_i,y_i)|}{N}}
\end{equation}

\begin{equation}
  \label{eq:linfnorm}
  L_\infty=\max_{\substack{1 \leq i \leq N}}|T_{cnn}(x_i,y_i)-T_{Nek5000}(x_i,y_i)|
\end{equation}

\begin{figure}[!htb]
%\begin{spacing}{1.0}
\centering
\hspace*{-1cm}
\includegraphics[scale=0.43,trim={10 0 30 10},clip]{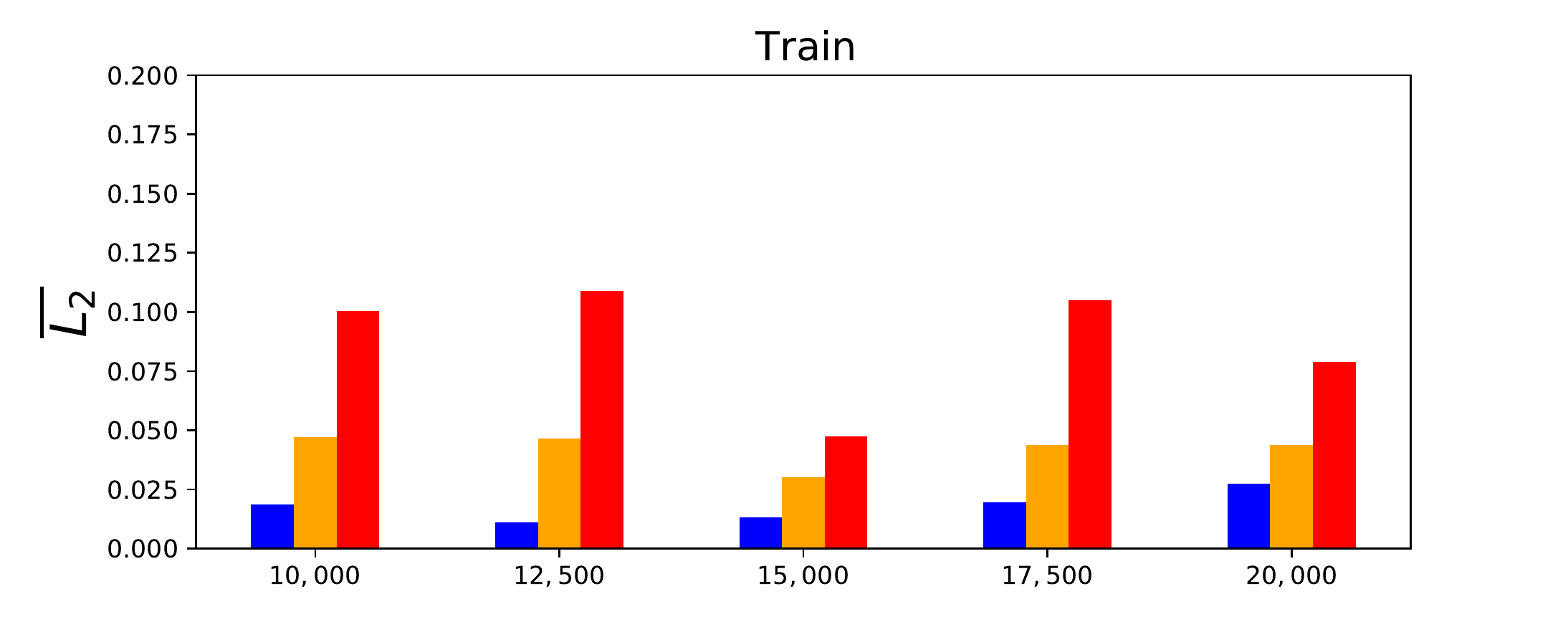}
\hspace*{0cm} 
\includegraphics[scale=0.43,trim={10 0 30 10},clip]{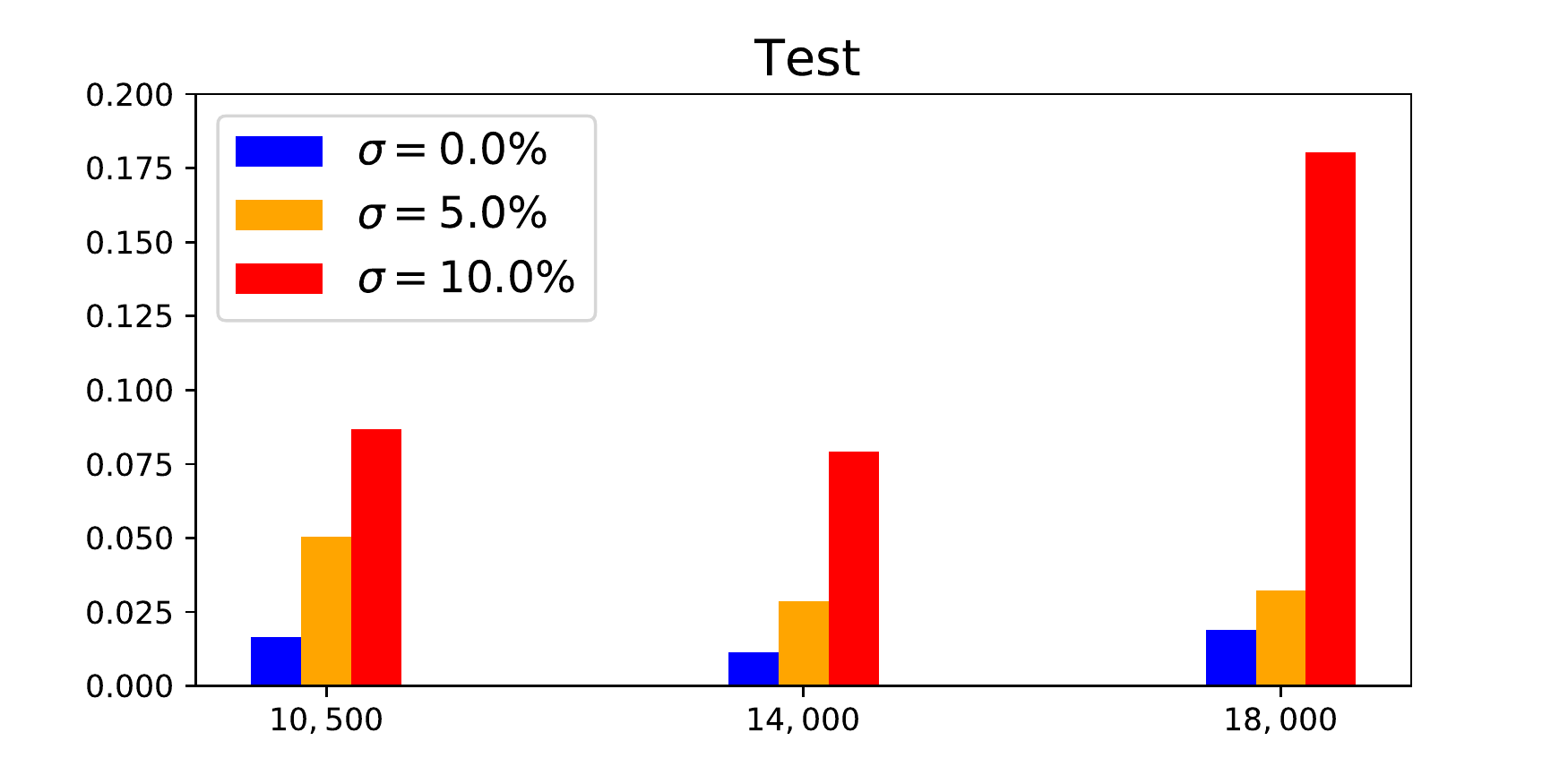}
\medskip
\hspace*{-1cm} 
\includegraphics[scale=0.43,trim={10 0 30 30},clip]{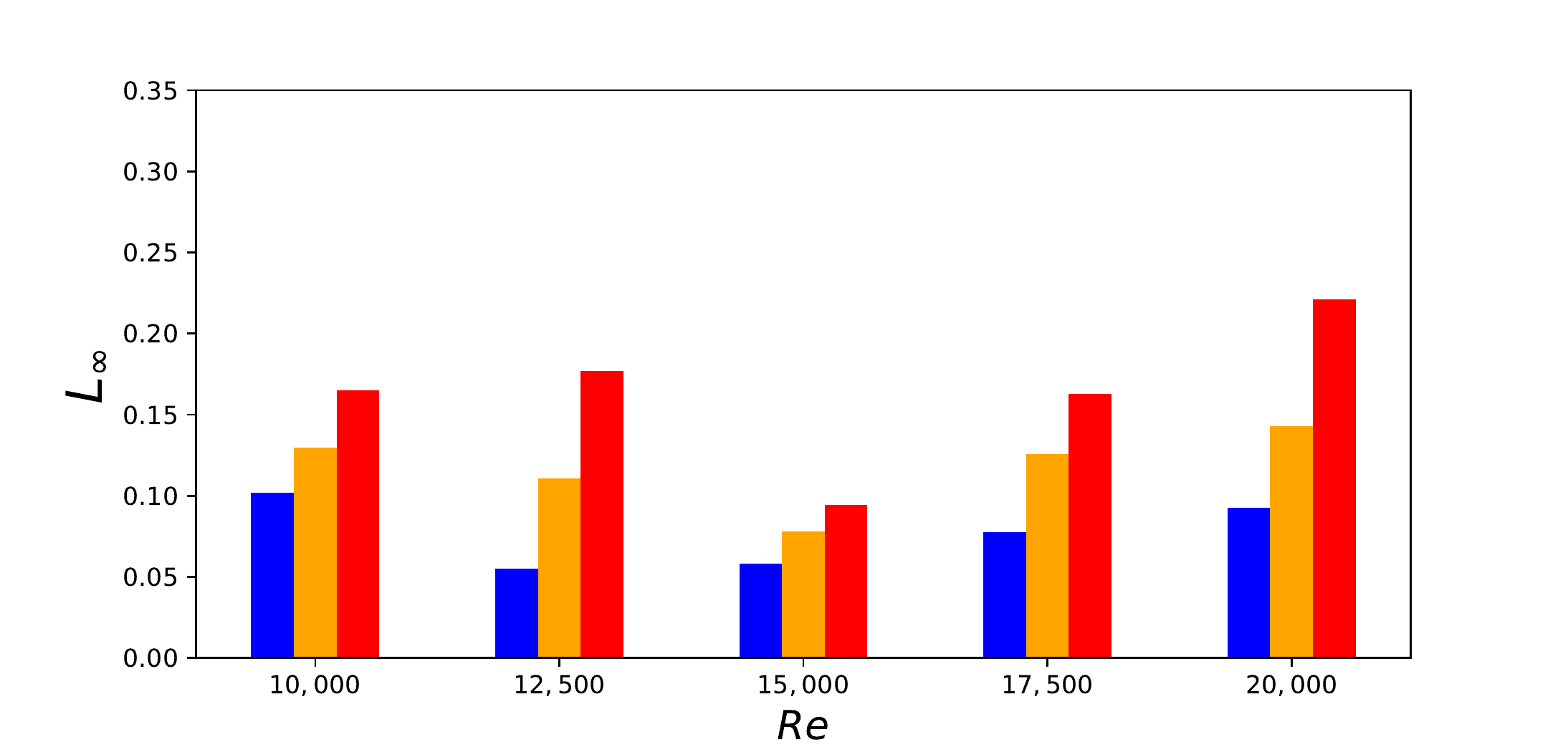}
\hspace*{0cm} 
\includegraphics[scale=0.43,trim={10 0 30 10},clip]{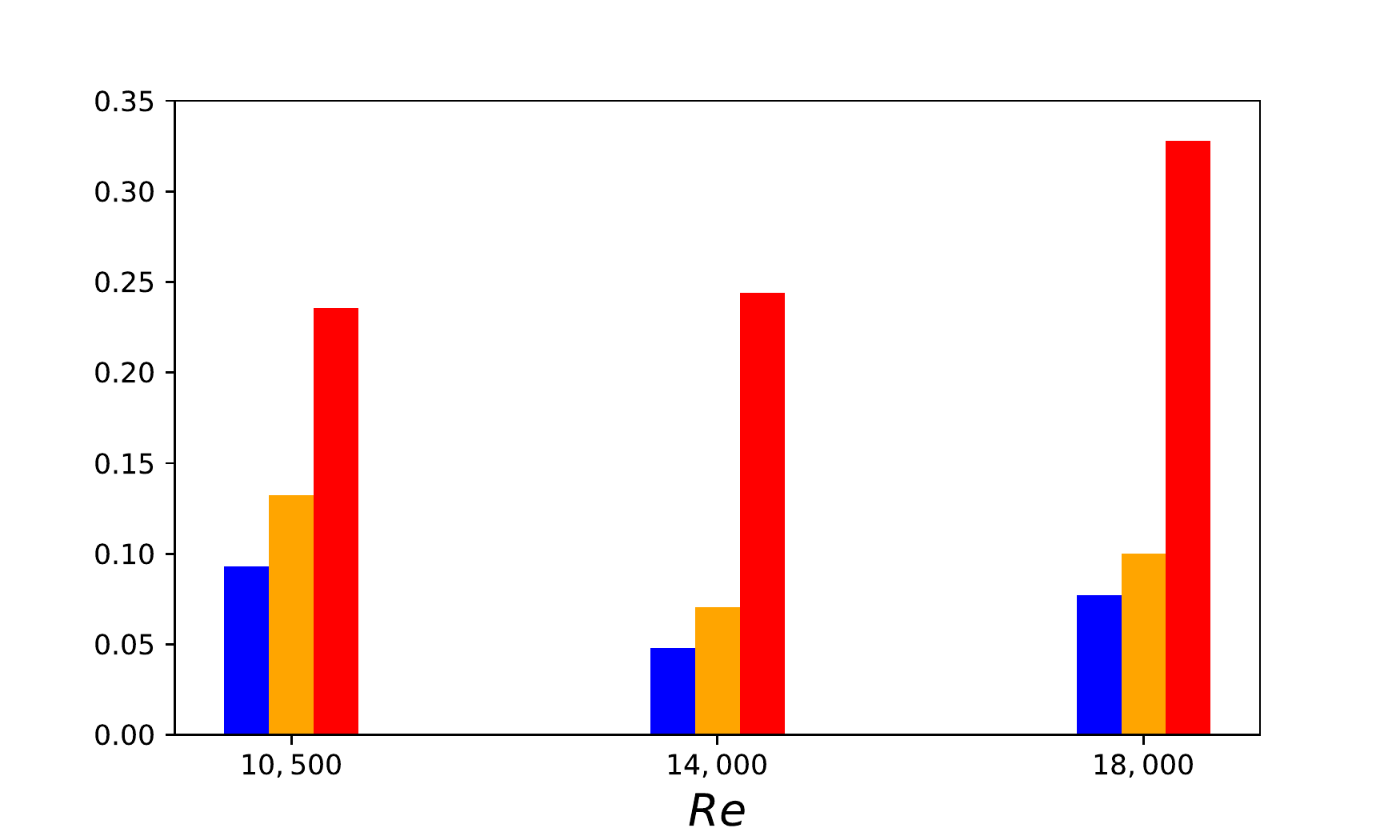}
\caption{Histogram of \(\overline{L_2}\) and \(L_{\infty}\) norms for conditions with high Prandtl numbers, \(Pr=10.7\).}
\label{fig:norms}
%\end{spacing}
\end{figure}

Fig.~\ref{fig:norms} provides histograms of these two types of errors in terms of dimensionless units considering both training and testing results. These results are provided only for the high Prandtl cases once this is the condition of more interest regarding MSR applications. This histograms also provides the results with the inclusion of randomness in the measurement points, i.e.  sensitivity analysis with \(\sigma=5\%\) and \(\sigma=10\%\).

%
%\vspace{8pt}
%\vspace{8pt}
%

The histograms shown in Fig.~\ref{fig:norms} assist on the assessment of the CNN. The following items explores some of the aspects of these results in order to understand them in more details. This should allow for a better comprehension of which conditions causes a better or poorer performance of the CNN. Moreover, the source of the severest errors is inspected, i.e. when \(\sigma>0\), and a suggestion of how to treat them is also provided.

\begin{itemize}

\item[i.] It is clear that the trend observed in the sensitivity analysis for \(\sigma\) is consistent with what expected, i.e. the higher this parameter is, the greater becomes the prediction errors. Furthermore, the \(L_\infty\) norms are always higher than the \(\overline{L_2}\) norms. The latter aspect should not always be truth, but it is also somewhat expected.

\item[ii.] Both \(\overline{L_2}\) and \(L_\infty\) norms from test conditions with \(\sigma=0\) are always within the range stipulated by their adjacent training conditions. The same kind of evaluation cannot be directly done for cases with \(\sigma>0\) once this results essentially depends on random parameters. Hence, for cases featuring \(\sigma>0\) a statistical collect including various predictions would be more appropriate.

\item[iii.] It is well known that neural networks suffers from a major drawback when performing predictions beyond the range of the original training data~\cite{leonard1992}. Furthermore,  they are also suspect to have local areas of poor fit even within this range,  e.g. at the limits of the training set. Fig.~\ref{fig:norms} obviously reflect this fact. For this fact, it should be kept in mind that the CNN must be applied for cases within the range of conditions from the training scope, otherwise the predictions will likely be arbitrary.

\item[iv.] By examining in more details the plots from Fig.~\ref{fig:noise_highPr}, we observe that poorer predictions when including random perturbations (especially with \(\sigma=10\%\)) are manifested in the form of bias rather than scattered errors. This is a good indication that such errors may be avoided with some kind of denoising techniques, here we provide.~\cite{vincent2010} as a reference describing a promising technique for this purpose.

\end{itemize}

\subsection{Demonstration use of the CNN for a MSR case}

This section shows a demonstration use of the CNN to predict the temperature field considering a MSR core cavity. For this purpose, a total of 18 measurement points were considered on the walls of the cavity, following what proposed in Fig.~\ref{fig:msr_cnn}. As a demonstration case, a single flow condition has been used for both training and predicting the dimensionless temperature field. The condition has been previously defined in Section~\ref{sec:demo_msr} and it has \(Re=18,000\) and \(Pr=10.7\).

\begin{figure}[htb]
    \centering % <-- added
\hspace{-1.5cm}
\begin{subfigure}{0.25\textwidth}
  \centering 
  \includegraphics[scale=0.5]{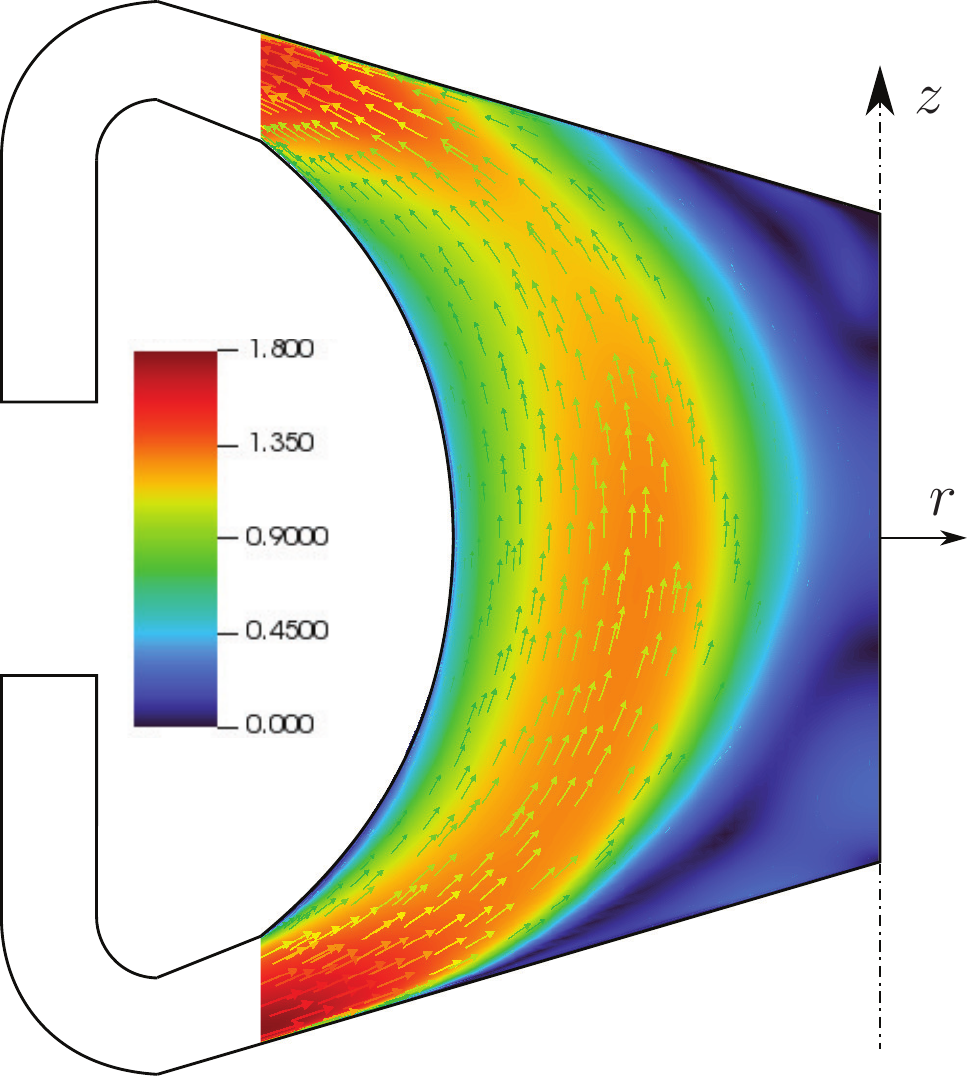}
  \caption{}
  \label{fig:nek_vel}
\end{subfigure}\hfil % <-- added
\hspace{-0cm}
\begin{subfigure}{0.25\textwidth}
 \centering 
 \includegraphics[scale=0.5]{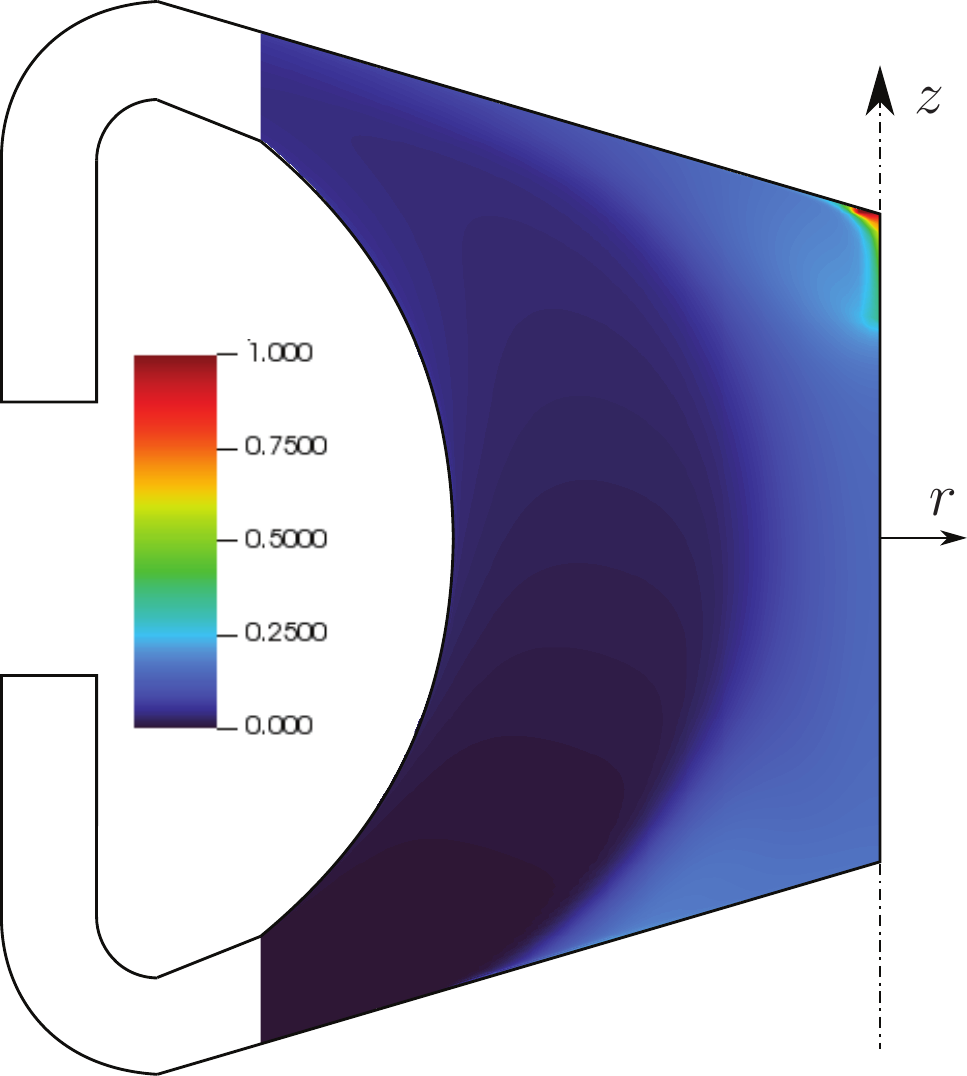}
 \caption{}
  \label{fig:cnn_temperature}
\end{subfigure}\hfil % <-- added
\hspace{-0cm}
\begin{subfigure}{0.25\textwidth}
 \centering 
 \includegraphics[scale=0.5]{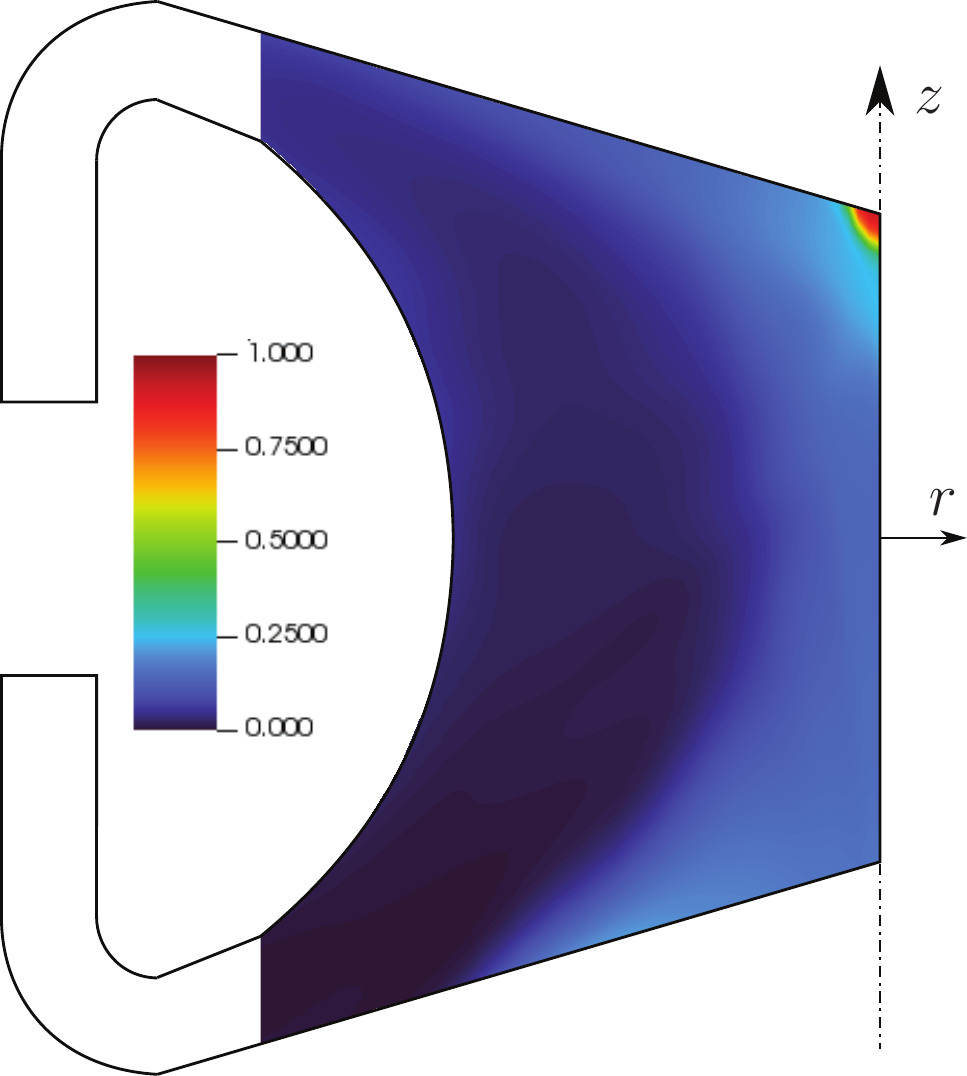}
 \caption{}
  \label{fig:cnn_predictions}
\end{subfigure}\hfil % <-- added
\caption{Physical field solutions: (a) velocity field from the Nek5000 simulation; the dimensionless temperature field from (b) Nek5000, and (c) the CNN model.}
\label{fig:msr_temperature}
\end{figure}

Fig.~\ref{fig:msr_temperature} presents the results of the MSR case considered including (a) the velocity field obtained from Nek5000 and (b) the temperature solution also obtained from Nek5000 and (c) the temperature field predicted by the CNN.

%\begin{table}[!htb]
%\caption{CNN accuracy for a MSR cavity.}
%\begin{center}
%\label{tab:gradp}
%\begin{tabular}{c c c}
%\hline
%$\overline{L_2}$ & $L_\infty$ & $\epsilon_{max}$  \\
%\hline
% 0.008  & 0.46 & 0.025  \\ 
%\hline
%\end{tabular}
%\end{center}
%\end{table}

\label{sec:MSR_results}

Fig.~\ref{fig:msr_temperature} clearly shows that the CNN was able to provide an accurate prediction for the temperature field inside the cavity of the MSR. Interestingly, the solution provided by the CNN is somehow better characterized by a diffusive phenomena when compared to the one obtained via CFD. This is particularly evident when observing the hot-spot region. Specifically at this location, the Nek5000 solution features a temperature gradient that is "pushed" in the streamwise direction towards the upper wall. Differently, the same behaviour is not observed in the CNN predicted field. Such difference in behaviours is potentially explained by the nature of the Helmholtz equation itself. Lastly, the predictions from the CNN resulted a \(\overline{L_2}=0.01\) and the maximum temperature with only \(3\%\) of relative error.

\section{Conclusions and Future Works}

The present work demonstrates the capabilities of a new physically-informed CNN model to predict and reconstruct the temperature field based on a limited set of measurements taken at the boundaries of a fluid domain. The CNN employed features a data-driven approach and it has been first proposed in Ref.~\cite{ponciroli2021convolutional}. The authors of this reference tested the CNN considering three-dimensional solution fields of the Laplace and the Helmholtz equations using the Boundary Element Method. In the present study, the CNN was successfully applied for reconstructing the temperature field of a heated channel driven by an incompressible fluid at various flow conditions. Furthermore, the CNN also proved to be successful for reconstructing the temperature field considering a much more complex configuration of a MSR.

The proposed methodology has demonstrated significant performance in all considered cases. The heated channel predictions for moderate Prandtl (\(Pr=1.0\)) and the MSR core-cavity demonstration case are especially noteworthy. The former should be underscored once the CNN successfully accounted for the non-linear behavior of turbulence transition as the train data set ranges from laminar, starting at \(Re=100\), up to turbulent at a higher Reynolds of \(Re=12,800\). The latter demonstrates how feasible it is to deploy the proposed methodology for ultimate application of a MSR core cavity.

Furthermore, the results have demonstrated that the CNN predictions are physically consistent. As described in Section~\ref{sec:cnn}, the Helmholtz equation is accounted in a physical layer of the CNN formulation. This is a key aspect in order to make the present technique superior to a general Machine Learning algorithm. Interestingly, the temperature fields reconstructed by this model are characterized by a more diffusive behavior when compared to the original data obtained via CFD analysis. 

For future works, other physical formulations might also be included besides the helmholtz-kirchhoff equation. In that sense, other fields, e.g. velocity, may also be assimilated by the CNN, which in principle could make the model even more efficient. Furthermore, the present CNN algorithm will be extended to the so-called Long-Short Term Memory (LSTM) model, featuring a Recurrent Neural Network (RNN). Unlike the standard CNN, this type of model has the ability to track long-term dependencies in the form of input sequences. This way, the method can deal with transients besides steady-state conditions. Finally, work is to be done on the reconstruction of three-dimensional temperature fields not only using CFD, but also considering real experiments.

\section{ACKNOWLEDGMENTS}

This research is being performed using funding received from the DOE Office of Nuclear Energy’s Nuclear
Energy University Program under contract DE-NE0008984.

Argonne National Laboratory’s work was supported by the U.S. Department of Energy, Office of Nuclear
Energy, under contract DE-AC02-06CH11357. This material is based upon work also supported by the
National Science Foundation Graduate Research Fellowship under Grant No. DGE 1256260.

\setlength{\baselineskip}{12pt}

%Bibliography
\bibliographystyle{ieeetr}  
\bibliography{templateArxiv}

\begin{thebibliography}{10}

\bibitem{godin1996}
O.~A. Godin, ``The kirchhoff–helmholtz integral theorem and related
  identities for waves in an inhomogeneous moving fluid,'' {\em The Journal of
  the Acoustical Society of America}, vol.~99, no.~4, pp.~2468--2500, 1996.

\bibitem{raissi2019}
M.~Raissi, P.~Perdikaris, and G.~Karniadakis, ``Physics-informed neural
  networks: A deep learning framework for solving forward and inverse problems
  involving nonlinear partial differential equations,'' {\em Journal of
  Computational Physics}, vol.~378, pp.~686--707, 2019.

\bibitem{ponciroli2021convolutional}
R.~Ponciroli, A.~Rovinelli, and L.~Ibarra, ``A convolutional neural
  network-based approach to field reconstruction,'' {\em arXiv},
  vol.~2108.13517v1, 2021.

\bibitem{durbin2014}
{Karthik Duraisamy and Paul Durbin}, ``Transition modeling using data driven
  approaches,'' Tech. Rep. Proceedings of the Summer Program, pp. 427, Center
  for Turbulence Research, Jan. 2014.

\bibitem{boiko1994}
A.~V. Boiko, K.~J.~A. Westin, B.~G.~B. Klingmann, V.~V. Kozlov, and P.~H.
  Alfredsson, ``Experiments in a boundary layer subjected to free stream
  turbulence. part 2. the role of ts-waves in the transition process,'' {\em
  Journal of Fluid Mechanics}, vol.~281, p.~219–245, 1994.

\bibitem{jiang2021}
C.~Jiang, R.~Vinuesa, R.~Chen, J.~Mi, S.~Laima, and H.~Li, ``An interpretable
  framework of data-driven turbulence modeling using deep neural networks,''
  {\em Physics of Fluids}, vol.~33, no.~5, p.~055133, 2021.

\bibitem{wang2018}
J.-L. Wu, H.~Xiao, and E.~Paterson, ``Physics-informed machine learning
  approach for augmenting turbulence models: A comprehensive framework,'' {\em
  Physical Review Fluids}, vol.~3, Jul 2018.

\bibitem{kutz2013}
I.~Bright, G.~Lin, and J.~N. Kutz, ``Compressive sensing based machine learning
  strategy for characterizing the flow around a cylinder with limited pressure
  measurements,'' {\em Physics of Fluids}, vol.~25, no.~12, p.~127102, 2013.

\bibitem{berkooz1993}
G.~Berkooz, P.~Holmes, and J.~L. Lumley, ``The proper orthogonal decomposition
  in the analysis of turbulent flows,'' {\em Annual Review of Fluid Mechanics},
  vol.~25, no.~1, pp.~539--575, 1993.

\bibitem{russel1975}
R.~D. Russell and L.~F. Shampine, ``Numerical methods for singular boundary
  value problems,'' {\em SIAM Journal on Numerical Analysis}, vol.~12, no.~1,
  pp.~13--36, 1975.

\bibitem{gulli2017}
A.~Gulli and S.~Pal, {\em Deep learning with Keras}.
\newblock Packt Publishing Ltd, 2017.

\bibitem{goodfellow2016}
I.~Goodfellow, Y.~Bengio, and A.~Courville, {\em Deep Learning}.
\newblock MIT Press, 2016.

\bibitem{kingma2015}
D.~P. Kingma and J.~Ba, ``Adam: A method for stochastic optimization,'' {\em
  arXiv}, vol.~abs/1412.6980, 2015.

\bibitem{bishop1995}
C.~M. Bishop, {\em Neural Networks for Pattern Recognition}.
\newblock Oxford University Press, Inc., 1995.

\bibitem{nek5000v19}
{Argonne National Laboratory, Chicago, Illinois}, ``Nek5000 version 19.0.''
  https://github.com/Nek5000/Nek5000, 2019-06-23.

\bibitem{patera1984}
A.~T. Patera, ``A spectral element method for fluid dynamics: Laminar flow in a
  channel expansion,'' {\em Journal of Computational Physics}, vol.~54, no.~3,
  pp.~468 -- 488, 1984.

\bibitem{wilcox1993}
D.~C. Wilcox, {\em Turbulence modelling for {CFD}}.
\newblock DCW Industries, La Ca{\~n}ada, 1993.

\bibitem{merzari2013}
E.~Merzari, W.~Pointer, and P.~Fischer, ``Numerical simulation and proper
  orthogonal decomposition of the flow in a counter-flow t-junction,'' {\em
  Journal of Fluids Engineering}, vol.~135, p.~091304, 09 2013.

\bibitem{shaver2020}
P.~Fischer, J.~Lottes, S.~Kerkemeier, O.~Marin, K.~Heisey, A.~Obabko,
  E.~Merzari, and Y.~Peet, ``{Nek5000 developments in support of industry and
  the NRC},'' Technical Report ANL/NSE-20/48, Argonne National Laboratory,
  2020.

\bibitem{beer2018}
G.~Beer, I.~Smith, and C.~Duenser, {\em The Boundary element method with
  programming : for engineers and scientists}.
\newblock Wien: Springer, 2008.

\bibitem{yamaji2014}
B.~Yamaji, A.~Aszódi, M.~Kovács, and G.~Csom, ``Thermal–hydraulic analyses
  and experimental modelling of msfr,'' {\em Annals of Nuclear Energy},
  vol.~64, pp.~457--471, 2014.

\bibitem{rouch2014}
H.~Rouch, O.~Geoffroy, P.~Rubiolo, A.~Laureau, M.~Brovchenko, D.~Heuer, and
  E.~Merle-Lucotte, ``Preliminary thermal–hydraulic core design of the molten
  salt fast reactor (msfr),'' {\em Annals of Nuclear Energy}, vol.~64,
  pp.~449--456, 2014.

\bibitem{leite2021}
{Leite, V. C., Fang, J., Reger, D., Merzari, E., Yuan, H., Shaver, D.},
  ``{Initial use of Nek5000/Cardinal to improve closure models in Pronghorn},''
  Tech. Rep. ANL/NSE-21/45, Argonne National Laboratory, 07 2021.

\bibitem{leonard1992}
J.~Leonard, M.~Kramer, and L.~Ungar, ``A neural network architecture that
  computes its own reliability,'' {\em Computers and Chemical Engineering},
  vol.~16, no.~9, pp.~819--835, 1992.
\newblock An International Journal of Computer Applications in Chemical
  Engineering.

\bibitem{vincent2010}
P.~Vincent, H.~Larochelle, I.~Lajoie, Y.~Bengio, and P.-A. Manzagol, ``Stacked
  denoising autoencoders: Learning useful representations in a deep network
  with a local denoising criterion,'' {\em Journal of Machine Learning
  Research}, vol.~11, no.~110, pp.~3371--3408, 2010.

\end{thebibliography}

\end{document}